\documentclass[nolinenumbers, twocolumn, twocolappendix, trackchanges]{aastex701}

\usepackage{graphicx} 
\usepackage{xspace}
\usepackage{framed} 
\usepackage{txfonts}
\usepackage{rotating}
\usepackage{ulem}
\usepackage{float}
\usepackage{enumitem}
\usepackage{amssymb}
\usepackage[dvipsnames]{xcolor}
\usepackage{sistyle}
\SIthousandsep{,}
\usepackage{natbib}

\newcommand{\fsf}{f_{\rm sf}}
\newcommand{\Nc}{N_{\rm c}}
\newcommand{\taustar}{\tau_{\rm *}}
\newcommand{\taup}{\tau_{\rm +}}
\newcommand{\taum}{\tau_{\rm -}}
\newcommand{\taudep}{\tau_{\rm dep}}

\newcommand{\Sgas}{\Sigma_{\rm gas}}
\newcommand{\Smol}{\Sigma_{\rm H_2}}

\newcommand{\SSFR}{\Sigma_{\rm SFR}}

\newcommand{\tff}{t_{\rm ff}}

\newcommand{\alphvir}{\alpha_{\rm vir}}
\newcommand{\eff}{\epsilon_{\rm ff}}

\begin{document}

\title{The Rhythm of the ISM: Tracing the Timescales of Gas Evolution and Star Formation across Galactic Environments}

\author[orcid=0009-0003-0415-404X,sname='Zuzanna Kocjan']{Zuzanna Kocjan}
\affiliation{Department of Astronomy, University of Maryland, College Park, MD 20742, USA}
\email[show]{zkocjan@umd.edu}  

\author[orcid=0000-0002-6648-7136,sname='Vadim Semenov']{Vadim A. Semenov}
\affiliation{Center for Astrophysics, Harvard \& Smithsonian, 60 Garden St, Cambridge, MA 02138, USA}
\email[show]{vadim.semenov@cfa.harvard.edu} 

\begin{abstract}

We investigate the physical origin of the star formation scaling relations between the gas depletion time, the star-forming gas mass fraction, and the gas surface density, $\Sgas$, on kiloparsec scales, all of which are the key ingredients of the observed Kennicutt--Schmidt relation. To elucidate these trends, we employ an analytical framework that explicitly connects these kiloparsec-scale properties to the timescales governing the rapid, continuous ISM gas cycle on the scales of individual star-forming regions, including the formation, dispersal, and local depletion of star-forming gas. Using a suite of idealized disk galaxy simulations spanning a range of environments from dwarf and Milky Way-mass systems to a gas-rich starburst analog, we measure the timescales of the gas cycle and relate them to the dynamical and turbulent properties of the interstellar medium (ISM). We find that star-forming regions form on a timescale close to the vertical turbulent crossing time of the galactic disk, $\sim$3--30 Myr, which decreases at higher $\Sgas$ due to the increase in turbulent velocities in the ISM and the decrease in the disk thickness. In contrast, the local star formation and dispersal of such gas are set by the local conditions. Specifically, the local depletion time, $\sim$200--2000 Myr, is decreasing at higher $\Sgas$, as star-forming gas becomes denser and more efficient in forming stars. The lifetime of such gas is very short, $\sim$0.4-1 Myr, and only weakly increases with $\Sgas$. Together, our results demonstrate how the star formation properties of galaxies on kiloparsec scales emerge directly from the interplay between the galaxy-scale dynamics, ISM turbulence, and the state of star-forming gas.

\end{abstract}


\section{Introduction} \label{sec:intro}

Despite abundant gas reservoirs, galaxies form stars inefficiently. The typical gas depletion times in star-forming galaxies are $\taudep \equiv M_{\rm gas}/\dot{M}_* \sim 1$--$3$ Gyr when considering molecular gas alone \citep[e.g.,][]{kennicutt89,kennicutt98,wong02,leroy13,bolatto17,ellison20,sun23,wong24}, and $\taudep \sim 2$--$10$ Gyr when neutral gas is included \citep[e.g.,][]{kennicutt89,kennicutt98,bigiel_star_2008,saintonge11,delosreyes19}. Such Gyr-long depletion times exceed by orders of magnitude all relevant dynamical times in the ISM, including freefall, turbulent crossing, and orbital times \citep[e.g.,][see \citealt{semenov_physical_2017} for a summary]{kennicutt98,wong02,leroy08,daddi10}. Star formation is likewise inefficient on the scales of individual star-forming regions, which deplete gas on timescales $\taustar \equiv M_{\rm gas,sf}/\dot{M}_{*,\rm sf}$ of hundreds of Myr, observationally often described via a low efficiency per freefall time, $\eff = \tff/\taustar \sim 1\%$ \citep[e.g.,][]{zuckerman-evans74,zuckerman-palmer74,krumholz07,evans09,evans14,heiderman10,lada10,utomo18,krumholz19}. 

Substantial progress has been made in understanding the inefficiency of star formation, with stellar feedback identified as a key regulatory mechanism across scales \citep[see, e.g.,][for reviews]{mckee07,naab17}. Feedback from young stars, including ionizing radiation, stellar winds, and supernovae, ionizes and disperses star-forming regions \citep[e.g.,][]{spitzer78,hopkins11,dale_ionization--induced_2012,lopez14,mcleod19,chevance_molecular_2020,grudic21}, drives turbulence and other forms of thermal and nonthermal support \citep[e.g.,][]{ostriker10,faucher-giguere_feedback-regulated_2013,egorov23}, affects galaxy morphologies by limiting gravitational fragmentation and regulating the angular momentum content \citep[e.g.,][]{okamoto05,zavala08,agertz15,agertz16}, launches galactic winds, and regulates gas cycling between the ISM, circumgalactic medium, and beyond \citep[e.g.,][]{bouche10,lilly13,tumlinson17,peroux20,pandya23,kocjan_hot_2024}. On small scales, while turbulence and magnetic support contribute to low $\eff$ \citep{krumholz05,padoan_star_2011,federrath12,padoan14}, local feedback processes such as ionizing heating and protostellar outflows further suppress star formation \citep[e.g.,][]{federrath15,grudic21,appel22}. Together, these processes establish a complex, multi-scale, multi-physics baryon cycle in which the interplay between gas flows, star formation, and stellar feedback maintains low star formation efficiencies from individual regions to entire galaxies.

A generic framework that describes this baryon cycle and explicitly links the relevant timescales was introduced by \citet{semenov_physical_2017}. In this framework, ISM gas continuously cycles between actively star-forming and inert, non-star-forming states under the influence of processes that lead to the formation of star-forming regions (e.g., gravity and cooling) and disperse them (e.g., stellar feedback, turbulence, and shear). Because these processes operate on the relatively short dynamical timescales of the ISM, the cycle itself is rapid. Star formation, however, remains globally slow as in each cycle, only a small fraction of the gas mass is converted into stars before the gas is rendered non-star-forming by feedback and dynamical effects. Consequently, on Gyr-long timescales, the ISM is vigorously ``boiling'', while its gas reservoir is depleted only gradually, analogous to the slow evaporation of water boiling in a kettle.

Apart from providing an intuitive explanation for why depletion times are long, the framework also explains the roles played by feedback, local, and galactic environments in this process. The galactic environment controls depletion times by determining how rapidly average ISM gas can become star-forming under the effects of gravity, cooling, and compression driven by disk instabilities and ISM turbulence, thereby setting the duration of one cycle. The local properties, such as the densities of star-forming regions and their turbulent state, determine how rapidly such regions convert gas into stars, setting local depletion times ($\taustar$ defined above). Stellar feedback, in turn, controls the depletion time by regulating the lifetime of such regions, thereby determining the total number of such cycles needed for gas depletion (see Section~\ref{sec:model_overview} for further details). To turn this framework into a predictive model, one needs a model for such timescales and their dependencies on the environment. Developing such a model is the focus of this paper.

Observationally, such dependencies manifest themselves in the relation between the depletion times and gas content on kiloparsec and larger scales, or the Kennicutt--Schmidt relation between the surface densities of SFR and gas \citep[KSR;][]{schmidt59,schmidt63,kennicutt89,kennicutt98,delosreyes19}. When only molecular ISM is considered, in normal (non-starburst) star-forming galaxies, the SFR scales almost linearly with the surface density of molecular gas: $\SSFR \propto \Sigma_{\rm H_2}$ \citep[e.g.,][]{wong02,bigiel_star_2008,leroy_molecular_2013,bolatto17,ellison20,sun23,villanueva24,wong24}, implying that the depletion time is near-constant, $\tau_{\rm dep,H_2} \equiv \Sigma_{\rm H_2}/\SSFR \approx {\rm const}$, although a weak dependence is possible depending on the assumptions about the CO-to-H$_2$ conversion factors, star formation indicators, and the details of the fitting procedure. For example, \citet{sun23} report a range of the molecular KSR slopes, $\SSFR \propto \Sigma_{\rm H_2}^\alpha$ with $\alpha \approx 0.9\text{--}1.2$, depending on the specific choices for these assumptions \citep[see also][]{ellison21,villanueva24,wong24}.

The above-mentioned gas cycling framework can explain this near-universality of $\tau_{\rm dep,H_2}$ as a result of feedback regulating the lifetimes of both star-forming and molecular gas (which are generally distinct ISM states), leading to the independence of $\tau_{\rm dep,H_2}$ of the kiloparsec environment \citep{semenov19}.
In contrast, when the neutral gas is accounted for or active starburst environments are considered, the KSR becomes non-linear, $\SSFR \propto \Sigma_{\rm gas}^\alpha$ with $\alpha \sim 1.4\text{--}1.5$ \citep[e.g.,][]{bigiel_star_2008,delosreyes19,kennicutt21}. Such superlinear dependence implies that the depletion time of the neutral phase, which is more representative of the total gas content, $\taudep = \Sgas/\SSFR$, does in fact exhibit systematic trends with kiloparsec-scale conditions, $\Sgas$. A model for the environmental dependence of the ISM gas cycle needs to explain such a dependence.

The origin of the KSR and the trends of depletion times with the environment have been extensively studied in theoretical and computational works. One class of ISM models links kiloparsec-scale depletion times to the dependence of the gas density probability distribution function (PDF) on $\Sgas$ \citep[e.g.,][]{elmegreen02,kravtsov03,gnedin_emergence_2014,renaud12,kraljic14,kraljic24}. This approach is conceptually analogous to models of individual star-forming regions, which attribute low local star formation efficiencies, $\eff$, and therefore long local depletion times, $\taustar$, to the small fraction of self-gravitating gas produced by supersonic turbulence \citep{krumholz05,padoan_star_2011,federrath12,padoan14}. By assuming that the density PDF depends on turbulent properties such as the Mach number or the virial parameter, these models predict how $\eff$ varies with environment. However, even on small scales, key aspects of this framework remain debated (e.g., the shape of the density PDF, the roles of magnetic fields and protostellar feedback, and the replenishment of self-gravitating gas), while extending it to galactic scales is further complicated by the multiphase nature of the ISM \citep[e.g.,][]{audit05,saury14,colman25} and the still debated relative roles of feedback and gravity in driving turbulence \citep[e.g.,][]{elmegreen04,krumholz_is_2016}.

Another class of models attributes the observed dependence of $\taudep$ on $\Sgas$ to a quasi-equilibrium state of marginally stable galactic disks \citep[e.g.,][]{ostriker10,ostriker_maximally_2011,faucher-giguere_feedback-regulated_2013,hayward17,ostriker_pressure-regulated_2022}. In these models, disks are regulated by a balance between the vertical gravitational weight of the ISM, which promotes gas compression and star formation, and opposing support processes, with the ISM turbulence driven by stellar feedback playing a central role. Stronger feedback increases turbulence, inflates the disk, and lowers the mean gas density, thereby reducing the star formation rate and, in turn, the driving of turbulence by feedback. This self-regulating loop maintains disks near marginal stability, with star formation rates, turbulent velocities, and disk scale heights set by the requirement to balance the ISM weight, thereby linking them to kiloparsec-scale conditions. In addition to stellar feedback, such models can also incorporate the effects of gas accretion and transport within disks \citep[e.g.,][]{krumholz_unified_2018,forbes_balance_2014}. While successful in describing normal star-forming disks, extending this framework to more extreme or out-of-equilibrium systems, such as high-redshift galaxies or mergers, remains challenging, particularly in the absence of well-defined disks.

Alternatively, some models describe the small-scale ISM gas cycle via the timescales of processes assumed to dominate the gas evolution, with global depletion times on kiloparsec and larger scales reflecting the environmental dependence of these timescales \citep[e.g.,][]{madore10,elmegreen15,elmegreen18,burkert_bathtub_2017}. A common assumption is that gas evolution is governed primarily by gravitational timescales: star-forming gas forms from the ambient ISM on freefall times at characteristic large-scale densities (e.g., midplane or volume-averaged), while star formation and gas lifetimes on small scales are set by freefall times at characteristic densities of dense actively star-forming gas. The latter is often motivated by the observed near-universality of the star formation efficiency per freefall time, $\eff$, with its low value typically treated as a model parameter. The resulting low mass fraction of star-forming gas and long global gas depletion times are then attributed to the large separation between freefall times at average ISM densities and those at high densities, associated with active star formation (see Section~\ref{sec:dis}). However, while gravity is undoubtedly central to star formation, it is unclear why freefall times alone should govern gas evolution in regimes where, e.g., turbulent support is equally important or even dominates over gravity, as suggested by the high virial parameters estimated for the observed actively star-forming regions \citep[$\alphvir \sim 1$\text{--}$10$;][]{leroy16,mivilledeschenes17,sun22}. This indicates that the small-scale ISM cycle requires a more detailed treatment than assumed in these analytic models.

A more comprehensive approach to the small-scale ISM gas cycle is provided by so-called effective ISM models \citep[e.g.,][]{yepes97,sh03,braun12a}. In these models, the evolution of the star-forming gas reservoir is represented by a system of ordinary differential equations (ODEs) describing mass and energy conservation, with key physical processes modeled by source, sink, and exchange terms, including cooling, heating, supernova-driven evaporation \citep{yepes97,sh03}, and turbulence \citep{braun12a}. Conceptually, this description is analogous to the gas cycling framework introduced above, as different sink and source terms in such models can be defined in terms of the characteristic timescales \citep[see][and Section~\ref{sec:model_summary} below]{semenov_physical_2017}.

Solving the resulting ODEs system, often under steady-state assumptions, yields predictions for the total star formation rate and effective ISM pressure as functions of global properties, such as the mean gas density and $\Sgas$, with free parameters calibrated against observations (e.g., the observed KSR). Owing to their numerical efficiency, such models became the backbone for many large-volume cosmological simulations \citep[see, e.g.,][for recent reviews]{vogelsberger20,crain23}. However, when applied to regimes beyond those used for calibration, such as the early Universe, their predictions can diverge significantly from high-resolution simulations, suggesting that key aspects of the ISM gas cycle are oversimplified or missing \citep[see, e.g.,][for a direct comparison]{semenov25a}. This motivates a detailed investigation of the physical drivers of small-scale gas evolution across diverse galactic environments.

In this paper, we apply the gas cycling framework to investigate how the timescales of the small-scale ISM cycle depend on the environment and how this dependence gives rise to the trends in gas depletion time on kiloparsec scales. Unlike the analytic and numerical models described above, this framework does not assume specific functional forms for the relevant timescales; instead, making predictions with this framework requires these timescales to be measured or modeled. Here, we analyze a suite of idealized galaxy simulations spanning a range of environments, including a disk dwarf galaxy similar to NGC\,300 or M33, a Milky Way-mass galaxy, and a gas-rich $z \sim 2$ starburst analog. We use passive gas tracer particles to measure the relevant timescales of formation and dispersal of star-forming gas, and demonstrate that our framework can explain the gas depletion times and star-forming mass fractions produced in our simulations. Based on these results, we build a simple model that connects the gas evolution timescales to the dynamical timescales in the ISM.

The paper is structured as follows. In Section \ref{sec:methods} we provide a brief overview of our analytical framework and a description of the simulations used in our work. In Section \ref{sec:results} we present our main results, highlighting how the measured timescales relate to the dynamical processes within the ISM. In Section \ref{sec:dis}, we discuss our results and compare them to the models outlined above. Section \ref{sec:sum} summarizes our conclusions.

\section{Methods} \label{sec:methods}

\subsection{Theoretical framework} \label{sec:model_overview}

In the following, we present a brief summary of the gas cycling framework developed by \citealt{semenov_physical_2017} that we adopt throughout our analysis. This framework describes the connection between depletion times and the timescales of small-scale ISM processes. It can be applied to galaxies as a whole, as well as individual regions, such as kiloparsec-scale ISM patches, as we do in this paper. 

In this framework, the total gas reservoir mass within a given ISM patch consists of gas in the star-forming and non-star-forming states, which can be expressed in terms of surface densities, $\Sgas = \Sigma_{\rm sf} + \Sigma_{\rm nsf}$, and the fraction of gas in the star-forming state, $\fsf = \Sigma_{\rm sf}/\Sgas$. Using the time-averaged star formation rate (SFR) surface density, ${\SSFR}$, the total gas depletion time of the patch is $\taudep = \Sgas/{\SSFR}$, while the average depletion time of star-forming gas itself is $\taustar = \Sigma_{\rm sf}/{\SSFR}$.

The rate at which the average, non-star-forming ISM gas becomes star-forming, $F_{+}$, is set by the supply timescale $\taup$, the average time gas spends in the non-star-forming state before becoming star-forming:

\begin{equation} \label{eq:f_p}
    F_{+} = \frac{\Sigma_{\rm nsf}}{\taup} = \Sgas \frac{1 - \fsf}{\taup} ,\,
\end{equation}
Similarly, the rate at which star-forming gas is rendered non-star-forming, $F_{-}$, is controlled by the removal timescale $\taum$, which represents the average lifetime of gas in the star-forming state before being dispersed by feedback or other dynamical processes, such as turbulence or shear:
 
\begin{equation} \label{eq:f_m}
    F_{-} = \frac{\Sigma_{\rm sf}}{\taum} = \Sgas \frac{\fsf}{\taum}.
\end{equation}

In steady state, the above rates are in balance, such that $F_{+} = F_{-} + \SSFR$.\footnote{Deviations from a steady state can be accounted for explicitly by including corresponding terms in the equation \citep[see][]{semenov_physical_2017}, which we do not do here as we apply this framework to isolated galaxy simulations close to a steady state.} By substituting Equations \ref{eq:f_p} and \ref{eq:f_m} into this condition, the equilibrium star-forming fraction becomes

\begin{equation} \label{eq:fsf}
    \fsf = \frac{\taum}{\taup + \taum}.
\end{equation}

Combining this expression with the definition of the local depletion time of star-forming gas, $\taustar = \Sigma_{\rm sf}/{\SSFR}$, the global depletion time, $\taudep = \Sgas/{\SSFR}$, can be expressed as
\begin{equation} \label{eq:taudep}
    \taudep = \taustar + N_{\rm c}\, \taup,
\end{equation}
where
\begin{equation} \label{eq:Nc}
    N_{\rm c} = \frac{\taustar}{\taum}
\end{equation}

\noindent
represents the number of cycles a parcel of gas typically undergoes while cycling in and out of the star-forming state before being converted into stars (see \citealt{semenov_physical_2017} for more details).

Equations~\ref{eq:fsf}--\ref{eq:Nc} relate the star-forming properties on kpc scale, $\fsf$ and $\taudep$, to the timescales of the ISM gas cycle on the scales of individual star-forming regions, $\taup$, $\taum$, and $\taustar$. These relations help elucidate several long-standing puzzling features of global star formation by plugging in typical values of the relevant timescales suggested by observations and simulations of galaxies. In particular, they explain why star formation is globally inefficient---i.e., why the global depletion time $\taudep$ is much longer than the dynamical timescales in the ISM, setting the duration of the gas cycle $\taup$---and why the global depletion times of gas ($\taudep \sim 1\text{--}10$ Gyr) are substantially longer than the local depletion times of actively star-forming gas ($\taustar \sim 300\text{--}500$ Myr). Star formation is globally inefficient ($\taudep \gg \taup$, with $\taup \sim 10\text{--}100$ Myr set by the characteristic dynamical timescales in the ISM), because on each cycle, only a small fraction of the star-forming gas is converted into stars, as the star-forming gas lifetime is much shorter than the time required for its depletion, $\taum/\taustar \ll 1$. This fraction is small because star-forming regions are both intrinsically inefficient in forming stars (i.e., $\taustar$ is long\footnote{E.g., as a result of only a small fraction of gas being gravitationally unstable in a cold supersonic medium; \citealt{krumholz05,padoan_star_2011,federrath12,padoan14}}) and are quickly dispersed by stellar feedback and dynamical processes ($\taum$ is short). Consequently, many cycles are required to deplete the gas reservoir, $N_{\rm c} \gg 1$. Moreover, the global depletion time $\taudep$ is also longer than $\taustar$ because a significant fraction of the cycle is spent in the non-star-forming state (the $N_{\rm c} \taup$ term in Equation~\ref{eq:taudep}). This leads to a low fraction of star-forming gas $\fsf$ due to $\taum \ll \taup$ (Equation~\ref{eq:fsf}).

In general, the timescales of the small-scale ISM processes, $\taup$, $\taum$, and $\taustar$, can depend on the kiloparsec environment, including the surface density of gas $\Sgas$, stars, metallicity, star formation activity, and ISM turbulence state, among others. In this paper, we focus on the dependence of these timescales on $\Sgas$, as the primary dependence in shaping the observed KSR (see the Introduction). Specifically, we use galaxy simulations outlined in the next section to determine the dependencies $\taup(\Sgas)$, $\taum(\Sgas)$, and $\taustar(\Sgas)$, and combine them with the above equations to understand the origin of $\taudep(\Sgas)$ and $\fsf(\Sgas)$ trends. Note that, as $\SSFR \equiv \Sgas/\taudep(\Sgas)$, this also accounts for the average dependence on the star formation rates and the UV field intensities associated with active star formation.

\begin{figure}
    \centering
    \includegraphics[width=0.49\textwidth]{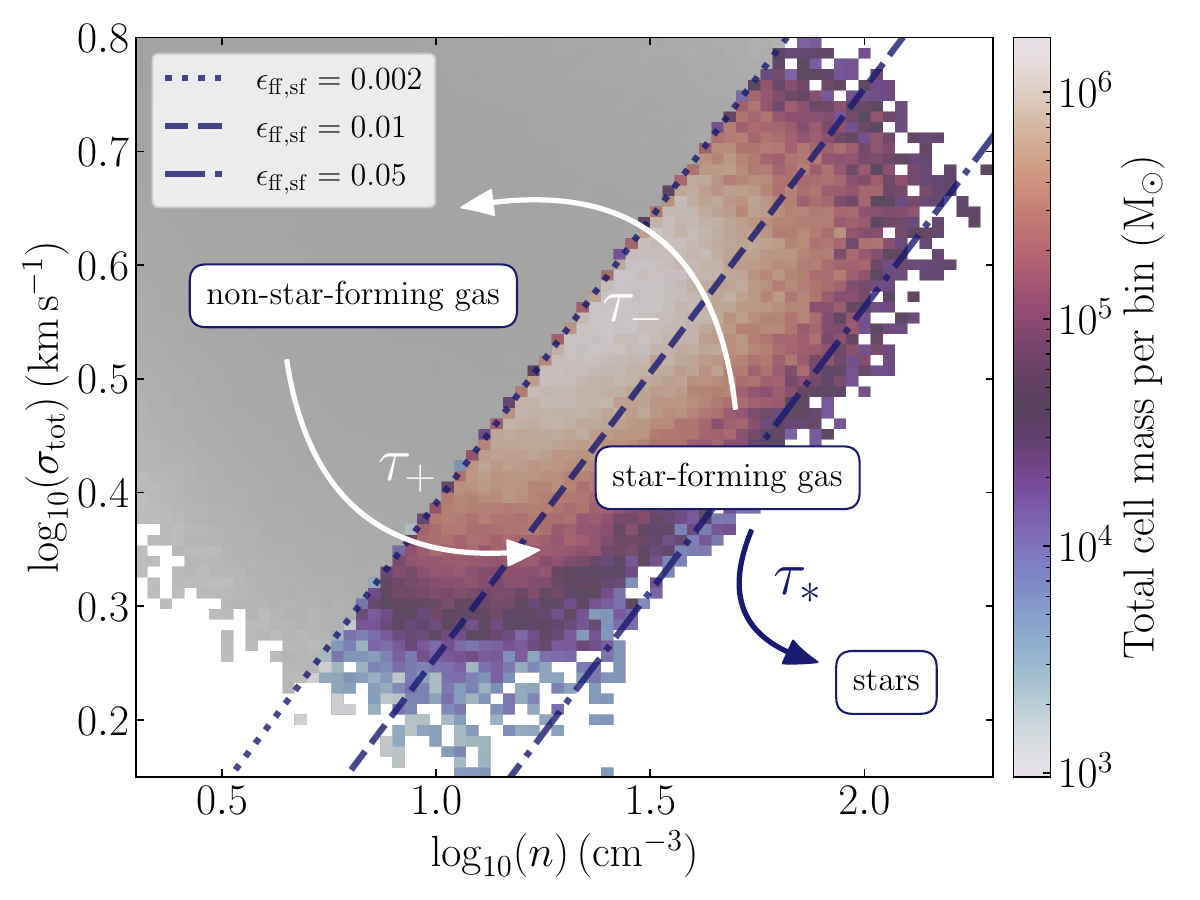}
    \caption{Illustration explaining the gas cycling in our model. The removal timescale $\taum$ represents the average lifetime of gas in the star-forming state before being dispersed by feedback or other dynamical processes. The supply timescale $\taup$ is the average time gas spends in the non-star-forming state before becoming star-forming. Once in the star-forming state, gas is converted into stars on a timescale $\taustar$.
    We select the star-forming gas as the gas with $\alphvir < \alpha_{\rm vir, sf} = 20$, which corresponds to $\eff > 0.002$ (dotted line; see the text). The dashed and dash-dotted lines show $\alphvir = 10.7$ and $\alphvir = 4.43$, respectively, which corresponds to $\eff = 0.01$ and $\eff = 0.05$ (see Equation~\ref{eq:epsff_P12}). These $\alphvir$ values roughly correspond to all, average, and the most active star formation. The environments in which these regimes occur are shown in Figure \ref{fig:eff_map} in purple, blue and orange, respectively. Importantly, the gas distribution in the $(n, \sigma_{\rm tot})$ plane determines how much gas lies below each of these $\alpha_{\rm vir}$ values. This distribution therefore affects the amount of gas in the star-forming state, the resulting star-formation efficiency, and ultimately the depletion time. As a consequence, the star-forming fraction and depletion time depend sensitively on the local environment.} 
    \label{fig:illust} 
\end{figure}

\begin{figure*}
    \centering
    \includegraphics[width=0.97\textwidth]{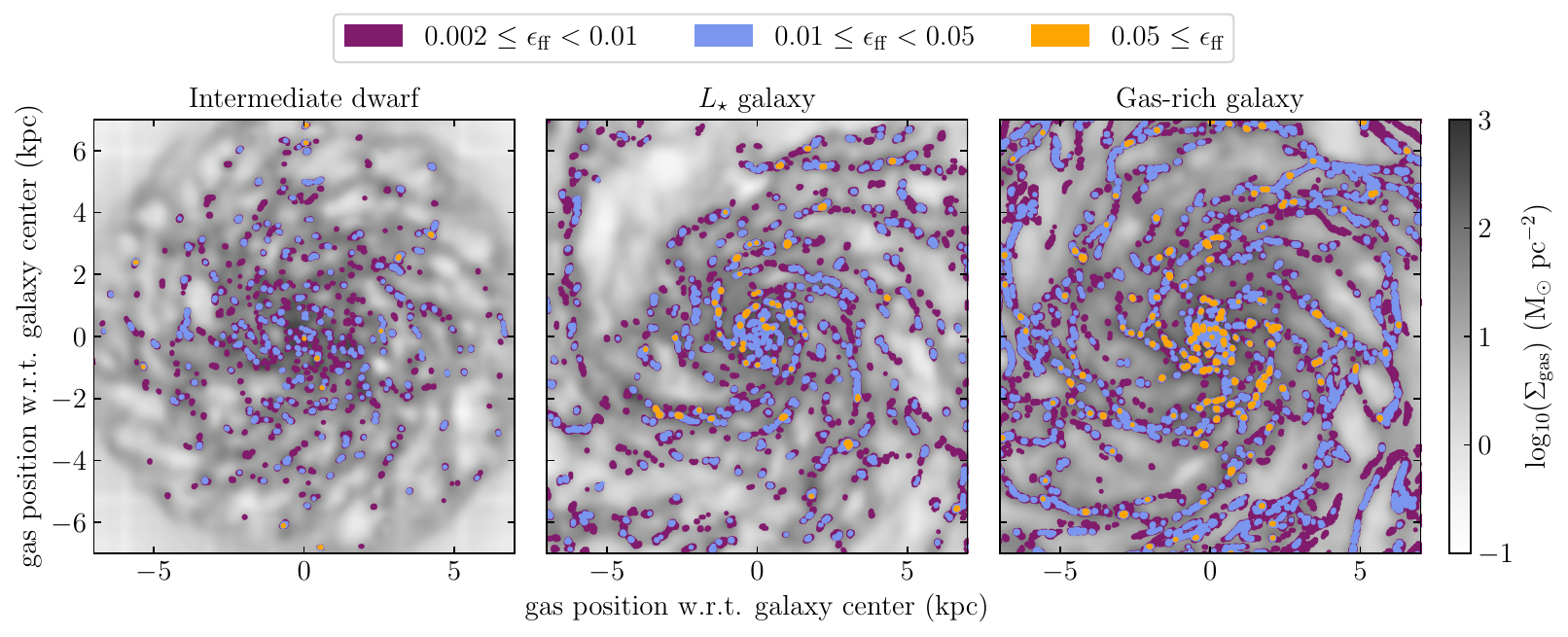}
    \caption{Slices of the three simulated galaxies used in this study, showing the star-forming regions identified with different $\eff$ values. Here $0.002 \le \eff < 0.01$ (purple), $0.01 \le \eff < 0.05$ (blue), and $\eff \ge 0.05$ (orange) correspond respectively to all, average, and the most active star formation. These selections are equivalent to applying virial-parameter cuts of $\alphvir < 20$, $10.7$, and $4.43$ (see Equation~\ref{eq:epsff_P12}). The contrast between the intermediate dwarf, $L_*$, and gas-rich galaxies (see Section \ref{sec:sims} for their descriptions) demonstrates that the fraction and spatial distribution of star-forming gas depend strongly on the kiloparsec-scale environment. For reference, the grayscale background shows $\Sgas$ smoothed on a scale smaller than 1 kpc for visual purposes, as it preserves some details of the underlying ISM structure.}
    \label{fig:eff_map}
\end{figure*}

\subsection{Simulation data} \label{sec:sims}

\begin{figure*}[t]
    \centering
    \makebox[\textwidth][c]{%
        \includegraphics[width=0.48\textwidth]{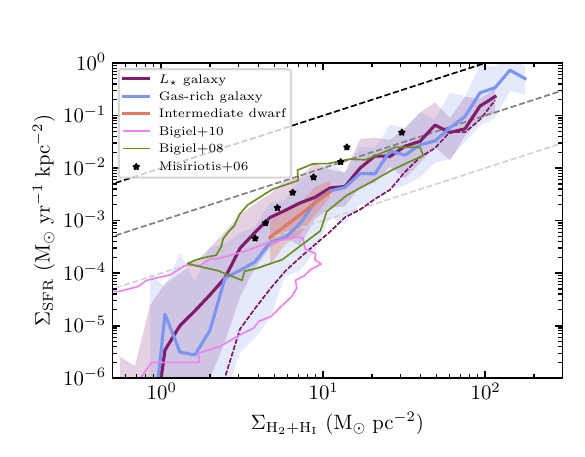}%
        \includegraphics[width=0.48\textwidth]{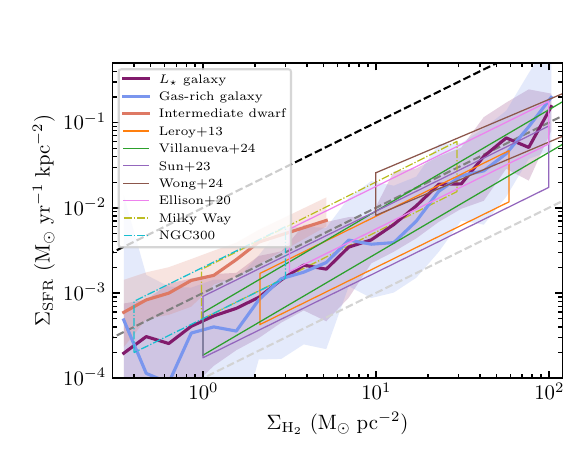}%
    }
    \caption{Relation between the star formation rate surface density, $\SSFR$, and the total (left panel) and molecular (right panel) gas surface densities in our simulated galaxies, compared to observational constraints. We show 1-kpc–patch–averaged median relations for the intermediate dwarf (orange), $L_\star$ galaxy (purple), and gas-rich galaxy (blue), with shaded regions indicating the 16--84th percentile scatter. To illustrate the effect of our H{\sc i} and H$_2$ selection, the purple dashed line in the left panel also shows the total gas KSR for the $L_\star$ galaxy, i.e., with $\Sgas$ instead of $\Sigma_{\rm H{\textsc i}+H_2}$. Dashed straight lines correspond to the constant depletion times $\taudep$ of 0.1 (black), 1 (gray), and 10 Gyr (light gray). In the left panel, our $\Sigma_{\rm H{\textsc i}+H_2}$–$\SSFR$ relations are compared to observations of nearby spiral galaxies from \citet{bigiel_star_2008, bigiel_extremely_2010}, and of the Milky Way from \citet{Misiriotis_2006}. In the right panel, we compare the $\Smol$–$\SSFR$ relations to observational measurements from \citet{leroy_molecular_2013, ellison20, sun23, villanueva24, wong24}. Additionally, the right panel includes a dash–dotted olive contour indicating the range $\tau_{\rm dep,H_2} \sim 0.5–2$~Gyr, estimated from radial profiles of $\SSFR$ and $\Smol$ compiled in Fig.~7 of \citet{Kennicutt_2012}, as well as a dash-dotted teal contour showing the corresponding range reported for NGC300 \citep{kruijssen_fast_2019}}
    \label{fig:combined_KSR}
\end{figure*}

We use three sets of isolated-galaxy simulations representing an intermediate dwarf (analogous to NGC300), an $L_{*}$ galaxy, and a gas-rich version of the $L_{*}$ galaxy, and therefore spanning a range of star-forming environments. These simulations are based on the runs used in \citealt{semenov_physical_2017, semenov_spatial_2021, semenov_cosmic-ray_2021}, respectively, except in all runs we unify the ISM, star-formation, and feedback model (summarized below) and adopt a uniform maximum spatial resolution of 20 pc for direct comparison. All simulations were carried out using the adaptive mesh refinement (AMR) N-body and gas hydrodynamics code ART \citep{kravtsov_adaptive_1997, rudd_effects_2008, gnedin_environmental_2011}, with on-the-fly modeling of UV radiation transfer (RT) coupled with gas thermochemistry \citep{gnedin14} and explicit treatment of turbulent energy on scales below the grid resolution (\citealt{semenov_nonuniversal_2016}, following the model of \citealt{schmidt_cosmological_2014}).

Each simulation is initialized with a dark matter halo, a gas disk and a stellar disk plus bulge component. The gas disk mass, scale radius and scale height are chosen to reproduce the intended morphological class. Below we summarize the properties of each model: \\

\textbf{Intermediate dwarf.} The NGC300 analogue is initialized with a Navarro–Frenk–White (NFW) profile, the total halo mass of $M_{200c} = 8.3\times10^{10}\,M_\odot$ (concentration of $c_{200c} \approx 15.4$) and an exponential gaseous disk with a radial scale length of $r_d \approx$ 1.39 kpc, a vertical scale height $h_{\rm d}$ of $\approx$ 280 pc and a total initial gas mass $M_{\rm gas} = 2.2\times10^{9}\,M_\odot$. The gaseous disk is initialized with a metallicity profile following the radial gradient of \citealt{2009_bersolin}: $Z(R)= (0.76\ Z_{\odot}) 10^{-0.077R}$, with $R$ in kiloparsecs. Further details about these initial conditions can be found in Section~2.2 of \citet{semenov_spatial_2021}. 

\textbf{\textit{L}$_*$ galaxy.} The $L_*$ isolated disk model is based on the AGORA code comparison project \citep{kim_agora_2016} initial conditions (see also \citealt{agertz_toward_2013}, \citealt{semenov_nonuniversal_2016}, \citealt{semenov_physical_2017}). The galaxy resides in a dark matter halo with $M_{200} \approx 1.1\times10^{12}\,M_\odot$ (concentration $c_{200} = 10$). It has an exponential stellar and gaseous disk with a radial scale length of $r_d \approx$ 3.4 kpc and a vertical scale height $h_{\rm d}$ of $\approx$ 340 pc. The gaseous disk contains a total gas mass of $M_{\rm gas} \approx 8.6\times10^{9}\,M_\odot$, corresponding to a gas fraction of $f_{\rm g} \approx 0.2$, and is initialized with a Milky Way--like metallicity gradient following \citealt{2018_stanghellini}, normalized such that $Z = Z_{\odot}$ at R = 8 kpc, yielding $Z(R)= (1.64\ Z_{\odot}) 10^{-0.027R}$, where $R$ is in kiloparsecs.

\textbf{Gas-rich galaxy.} 
The gas-rich model is a high-gas-fraction variant of this same AGORA $L_*$ setup. It retains the same radial scale length, vertical scale height, halo mass, concentration, and metallicity gradient as the fiducial model, but adopts an increased disk gas fraction of $f_{\rm g} \approx 0.4$. To mitigate the violent gravitational instability of such a gas-rich disk, we gradually increase the gas mass fraction, starting from simulations with a 20\% gas fraction at 600 Myr, and increasing it to 40\% by 650 Myr. While done manually, this approach can also mimic the rapid gas accretion seen in high-redshift galaxies. For more details, see section 3.3 in \citealt{semenov_cosmic-ray_2021}.

The modeling of the small-scale ISM processes, including heating and cooling, star formation and feedback, follows that in \citet{semenov_spatial_2021}. This includes the explicit on-the-fly transfer of the UV radiation field modeled using the Optically Thin Variable Eddington Tensor approximation \citep[OTVET;][]{gnedin01,gnedin14}, coupled with \citet{gnedin12} metallicity- and incident radiation-dependent cooling and heating model. The feedback from young stars is modeled by sourcing UV radiation that is self-consistently propagated by the RT solver and via injection of thermal energy and radial momentum in the amounts dependent on local gas density, metallicity, and cell size, following the results from simulations of SN remnants expanding into a nonuniform ISM by \citet{martizzi15}, with an additional momentum boost by a factor of 5 to account for the effects of SN clustering and cosmic-ray pressure.

The star formation rate in each cell is parameterized via the star formation efficiency per freefall time, $\tff = \sqrt{3\pi/32G\rho}$, as 

\begin{equation} \label{eq:rho_sfr}
    \dot{\rho}_\star = \frac{\eff \rho}{\tff}.
\end{equation}
In our simulations, the value of $\eff$ is dynamically modeled in each cell using the results of high-resolution MHD simulations of star-forming ISM from \citet{padoan_simple_2012}: 
\begin{equation} \label{eq:epsff_P12}
\eff = 0.9\, \rm exp{(-\sqrt{(\alphvir/0.53)})}, 
\end{equation}
where $\alphvir$ is the virial parameter of the gas on subgrid scale, which, for consistency with \citet{padoan_simple_2012}, is defined as that for a box with a side $\Delta$ as for a uniform sphere of radius $R/2$, following \citet{bertoldi-1992}:

\begin{equation} \label{eq:alph}
    \alpha_{\rm vir} \equiv \frac{5 \sigma^2_{1D}R}{GM} \approx 9.35 \frac{(\sigma_{\rm tot}/10\, \rm km\, s^{-1})^2 }{(n/100\, \rm cm^{-3})(\Delta\,/40 pc)^2}\,.
\end{equation}

Here the total velocity dispersion is defined as

\begin{equation} \label{eq:sigmatot}
    \sigma_{\rm tot} = \sqrt{\sigma_{\rm sgs}^2 + \sigma_{\rm thermal}^2},
\end{equation}

\noindent
where $\sigma_{\rm sgs}$ denotes the subgrid-scale turbulent component---dynamically modeled in each cell via tracking the cascade of turbulent energy from resolved scales, turbulent diffusion, and dissipation into heat \citep{semenov_nonuniversal_2016}---and $\sigma_{\rm thermal}$, the thermal (speed of sound) contribution. The thermal term is generally subdominant for the cold, star-forming gas, due to low local gas temperature.

Each galaxy model is evolved until the disk reaches a quasi‐steady state (typically several hundred Myr) and then sampled at regular intervals (every 20 Myr). Throughout the analysis, we calculate the timescales and other relevant quantities as detailed in the next section over patches with a linear size of 1 kpc. For each patch, the star formation rate surface density, $\SSFR$, is computed as the sum of the instantaneous star formation rates of all cells contained in the patch, divided by the patch area (1 kpc$^2$). We tested the sensitivity of our results to the patch size and found that the results remain unchanged when we vary it from 0.5 to 2 kpc.

\subsection{The ISM Gas Cycle in the Simulations} \label{sec:sims-cycle}

To put the framework introduced in Section~\ref{sec:model_overview} in the context of our simulations, Figure \ref{fig:illust} schematically illustrates the gas cycling and provides a visualization of the relevant timescales. The figure shows the distribution of the total velocity dispersion $\sigma_{\rm tot}$ and gas density $n$ in the NGC300 analogue, with star-forming gas occupying the colored region and non-star-forming gas shown in gray. As the local star formation efficiency in our simulations exponentially depends on $\alphvir$ (Equation~\ref{eq:epsff_P12}), and the value of $\alphvir$ changes exponentially between the warm, diffuse, subsonic ($\alphvir > 10^3$) and the cold, dense, supersonic ISM states ($\alphvir \lesssim 20$), the boundary between star-forming and non-star-forming states can be defined based on the $\alphvir$ value. In this work, we define star-forming gas as gas with $\alphvir < \alpha_{\rm vir,sf} = 20$, which corresponds to star-formation efficiency per free-fall time $\epsilon_{\rm ff} > 0.002$ and includes $>95\%$ of all star-forming gas in our simulations.\footnote{Note that this definition differs from $\alphvir = 10$ used in \citet{semenov_physical_2017}. This is because \citet{semenov_physical_2017} used fixed $\eff = 1\%$ together with this $\alphvir$ value to approximate Equation~\ref{eq:epsff_P12}. Here, we allow $\eff$ to vary continuously with $\alphvir$ and therefore adopt a higher $\alphvir$ to include the gas with low, but still sizeable star-formation efficiency between $\epsilon_{\rm ff} \approx 0.002\text{--}0.01$.}
The dash-dotted and dotted lines are indicating $\alphvir = 10.7$ and $\alphvir = 4.43$, corresponding to $\eff = 0.01$ and $\eff = 0.05$, respectively. Thus, the diagonal lines in the plot show, from top to bottom, all, average, and the most active star formation. The arrows indicate how gas circulates between phases, with the key timescales introduced in Section~\ref{sec:model_overview} also annotated. 

In the analysis presented below, we calculate the supply and removal timescales $\taup$ and $\taum$ using their definitions (Equations~\ref{eq:f_p} and \ref{eq:f_m}) with the fluxes in and out of the star-forming state calculated using passive Monte Carlo tracer particles that follow the gas flow in a Lagrangian fashion by stochastically sampling mass fluxes between cells \citep{genel-2013,semenov-2018}. Specifically, we compute $\taup$ from the ratio of the number of tracers that transition from the non-star-forming to the star-forming state to the number of tracers that are initially non-star-forming, while $\taum$ is obtained from the ratio of tracers that move from the star-forming to the non-star-forming state to those that are in the star-forming state initially. These ratios are then divided by the time interval between consecutive snapshots and inverted to obtain the corresponding transition timescales. 
The local depletion times, $\taustar$, are calculated from the local SFR density used in the subgrid model for star formation (Equation~\ref{eq:rho_sfr}), which, for consistency with this equation, were inversely averaged and weighted by mass over the star-forming cells in each kpc$\times$kpc patch: $\taustar = \langle (\tff/\eff)^{-1} \rangle^{-1}_{\rm sf}$.

Variations in galactic conditions modify the supply and removal timescales, $\taup$ and $\taum$, which regulate how gas moves through the $(n, \sigma_{\rm tot})$ space in Figure \ref{fig:illust}. As a result, changes in $\taum$ and $\taup$ reshape the distribution of star-formation efficiencies per free-fall time, $\epsilon_{\rm ff}$. To illustrate this, in Figure~\ref{fig:eff_map} we show the spatial distribution of $\eff$ in our simulations. The regions shown in purple, blue, and orange correspond to all, average, and the most active star formation, similarly to the dotted, dashed, and dash-dotted lines in Figure \ref{fig:illust}. The distribution of gas in the $(n, \sigma_{\rm tot})$ plane determines how much material lies below each of these $\alphvir$ (or $\eff$) lines. As a result, the fraction of gas that is star-forming, and therefore the resulting star-formation efficiency and depletion time, are not fixed quantities but depend sensitively on the kpc-scale environment. 

This environmental variation in the star formation activity manifests itself in the Kennicutt--Schmidt type relations between the surface densities of SFR and gas, which are shown in Figure~\ref{fig:combined_KSR}. 
To calculate the H$_2$ abundances in each patch, we used the $F_{\rm H_2} = \Sigma_{\rm H_2}/\Sigma_{\rm H}$ model of \citet{polzin24-h2} (see their Section 3.2), using the density-weighted metallicities and free-space UV radiation field strengths self-consistently followed in the simulation. Although our simulations include a molecular chemistry network, we find that reducing the resolution by a factor of 2 compared to \citet{semenov_spatial_2021} results in H$_2$ abundances reduced by a factor of $\sim 2$, as the local densities reached by the gas become lower. This effect can be corrected by properly adjusting the subgrid clumping factor of H$_2$ \citep[see][for the definition]{gnedin_environmental_2011}, or tying it directly to the Mach number of subgrid turbulence predicted in our simulations. We leave a detailed study of this effect to future work, and for the comparison with observations here, we use the \citet{polzin24-h2} model, which was calibrated against the higher-resolution simulations that reproduce the observed H$_2$ abundances. Note that this does not affect our results presented below, as star formation in our simulations is independent of molecular gas modeling. 

Following \citet{semenov_physical_2017}, we have also excluded cold H{\sc i} gas with $T < 1000$ K, which is expected to be in the optically thick cold neutral medium (CNM) and would therefore be missing in the observational measurements shown in the figure. The specific value of the temperature threshold is motivated by the requirement that the resulting $\Sigma_{\rm H{\textsc i}}$ does not extend above the typical observed upper bound of $\sim 10\ {\rm km\,s^{-1}}$ \citep[e.g.,][]{bigiel_star_2008}.
The horizontal deviation of the solid purple line from the dashed one at low $\Sigma_{\rm H{\textsc i}+H_2}$ indicates that the gas surface density is dominated by the molecular gas at $\Sigma_{\rm H{\textsc i}+H_2} \gtrsim 50\ {\rm M_\odot\,pc^{-2}}$, while at lower surface densities, a significant gas mass fraction ($\sim 50\%\text{--}70\%$) is in the optically thick CNM and ionized states. This is higher than the current estimates of the CNM mass fraction in environments similar to those in our simulations \citep[$\sim 30\%$; e.g.,][]{murray18,koch21}, although this value can be underestimated due to the uncertainty in the CNM structure. The solid and dashed lines in the figure, therefore, indicate the plausible range of KSR for different assumptions about the CNM and ionized gas mass contribution.

As described in the introduction, the KSR for molecular gas (the right panel of Figure~\ref{fig:combined_KSR}) is near-linear, which can be explained by the efficient regulation of molecular gas depletion times by feedback that limits the lifetimes of both star-forming and molecular gas, making $\tau_{\rm dep,H_2}$ near independent of the environment \citep[see][for a detailed study of the origin of the near-linear slope]{semenov19}. The slope in our simulations is linear at $\Sigma_{\rm H_2} < 10\ {\rm M_\odot\,pc^{-2}}$, and slightly steepens at higher surface densities. As a result, while the normalization of molecular KSR in our simulations is consistent with the reported observed values, its slope at high $\Sigma_{\rm H_2}$ is closer to the moderately super-linear estimates (e.g., \citealt{sun23} results for \citealt{bolatto13} and \citealt{gong20} $\alpha_{\rm CO}$ models; \citealt{villanueva24}).

As the left panel shows, the inclusion of atomic gas results in a more complex, non-linear dependence of $\SSFR$ (and therefore $\taudep$) on the gas surface density, with the strong steepening at low $\Sigma$ driven by the H{\sc i}-to-H$_2$ transition \citep[see, e.g.,][]{bigiel_star_2008,gnedin_environmental_2011}. This is because the sum of atomic and molecular ISM more closely represents the total available gas reservoir. For comparison, the dashed purple line in the left panel shows the KSR that includes all available gas in each patch (including ionized and optically thick H{\sc i}). The resulting KSR becomes even steeper due to the increasing contribution of gas components not included in H{\sc i}+H$_2$ at low surface densities. 

Accurately modeling the KSR shape, therefore, requires modeling both the dependence of the total gas depletion time, setting $\SSFR = \Sgas/\taudep$, and the transitions between ionized, atomic, and molecular phases, setting the gas mass fractions in these phases. In this paper, we focus on the first part and develop a model for the total gas depletion time as a function of $\Sgas$. Figure~\ref{fig:combined_KSR} shows that the KSR produced in our galaxy simulations agrees with observations reasonably well, supporting the use of these simulations to develop such a model in the next section. A model describing atomic and molecular mass fractions can either be calibrated directly against simulations \citep[see, e.g.,][]{polzin24-h2}, or parametrized via the characteristic timescales of H{\sc i} and H$_2$ formation and dispersal, analogously to the analysis presented below.

\section{Results} \label{sec:results}

\subsection{Overview of the theoretical model}

\begin{figure*}
    \centering
    \includegraphics[width=1.0\textwidth]{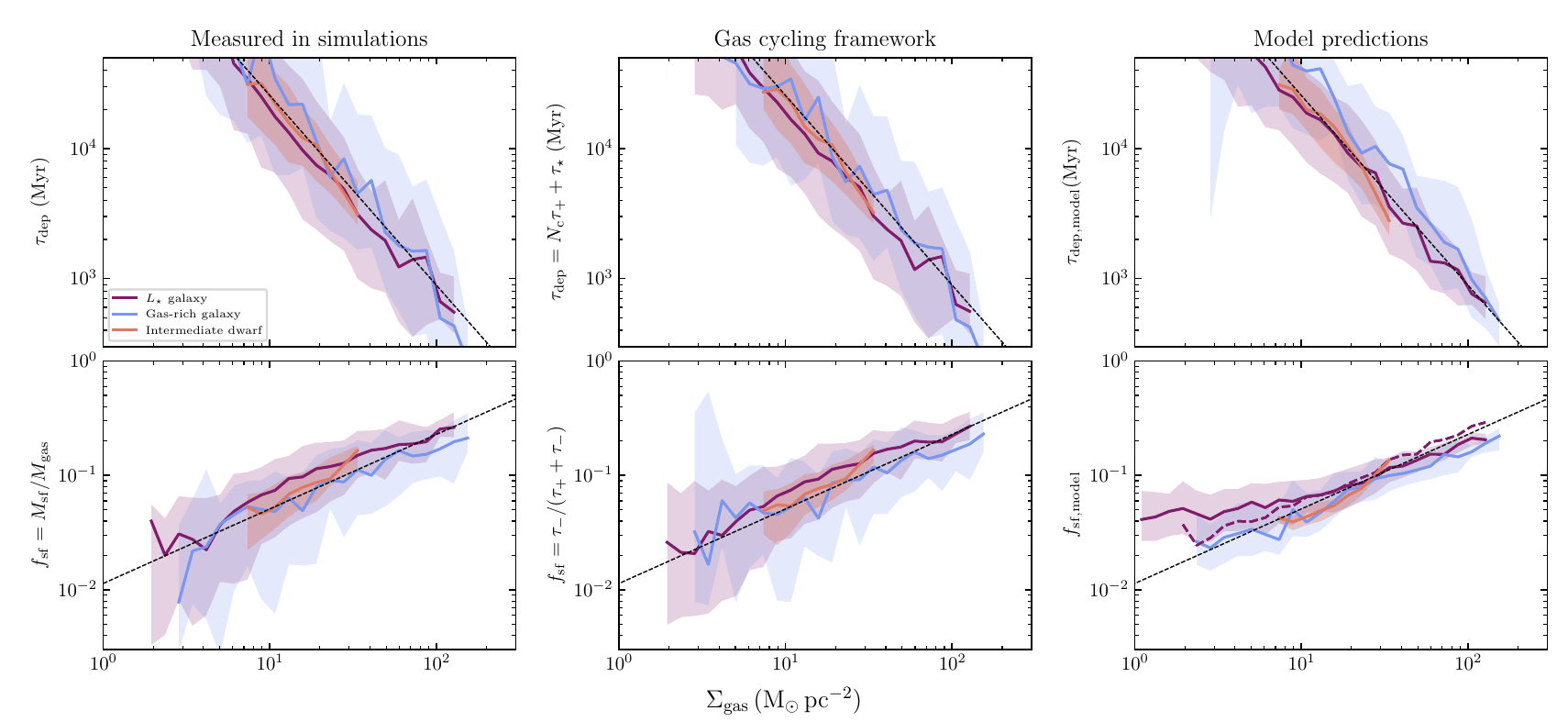}
    \caption{Comparison between the relations measured directly from the three simulations (left column) with those predicted using our gas-cycling framework outlined in Section~\ref{sec:model_overview} (middle column) and by our simple model summarized in Section~\ref{sec:model_summary} (right columns). Shown are the global depletion time, $\tau_{\rm dep}$ (top row), and the star-forming gas fraction, $f_{\rm sf}$ (bottom row), both as functions of the gas surface density, $\Sigma_{\rm gas}$. In the middle column, we show the predictions of our framework obtained by explicitly measuring the characteristic timescales of the gas cycle ($\taup$, $\taum$, and $\taustar$) using passive gas tracer particles and substituting them into Equations~\ref{eq:fsf}--\ref{eq:Nc}. The right column shows model predictions based on simple, physically motivated prescriptions for $\taup$, $\taum$, and $\taustar$, which we derive in Sections~\ref{sec:tau_plus}--\ref{sec:tau_minus}. Solid lines show the 1 kpc patch medians for the intermediate dwarf (orange), $L_*$ galaxy (purple), and gas-rich galaxy (blue), while the purple dashed line in the bottom right panel shows a prediction with $\taum$ calculated with tracer particles for the $L_*$ galaxy (see the text for details). For easier visual comparison, the black dashed lines show power-law fits to the measured medians, $\taudep \propto {\Sigma_{\rm gas}}^{-1.46}$ and $f_{\rm sf} \propto {\Sigma_{\rm gas}}^{0.65}$. Shaded regions indicate the 16--84th percentile scatter. The close match between the measured values and theoretical predictions demonstrates that the model successfully reproduces the dependence of star formation properties on gas surface density across a wide range of galactic environments.} 
    \label{fig:model_all} 
\end{figure*}

\begin{figure}
    \centering
    \includegraphics[trim = 3mm 0mm 0mm 6mm, clip, width=0.4\textwidth]{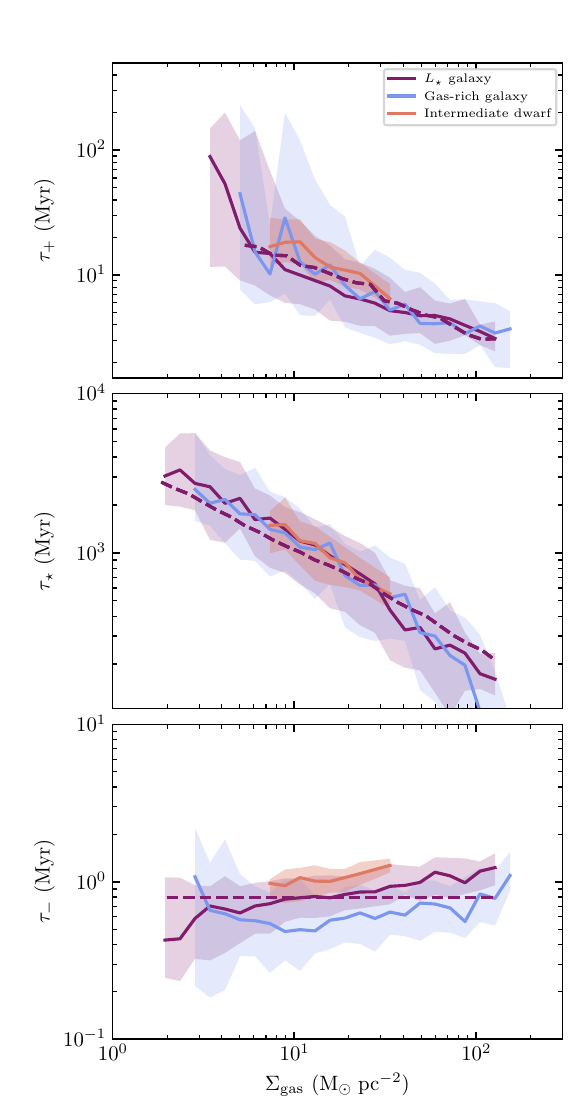}
    \caption{Dependence of the characteristic model timescales $\taup$, $\taustar$, and $\taum$ on gas surface density, $\Sigma_{\rm gas}$. The solid lines show the median values as measured in the simulation, while the purple dashed line shows predictions from the simple model, for the $L_*$ galaxy only (summarized in Section~\ref{sec:model_summary}). As $\Sigma_{\rm gas}$ increases, gas enters denser star-forming regions, resulting in shorter average depletion and supply timescales, $\taustar$ and $\taup$. This trend reflects the higher densities and more efficient compression of gas into the star-forming state at high $\Sigma_{\rm gas}$. In contrast, the removal timescale $\taum$ increases with $\Sigma_{\rm gas}$, as denser gas is more resistant to dispersal by feedback or turbulent motions.} 
    \label{fig:model_timescales} 
\end{figure}

Figure \ref{fig:model_all} provides an overview of how our gas-cycling framework (Section~\ref{sec:model_overview}) connects the key star-forming properties on a kiloparsec scale, i.e., the depletion time $\tau_{\rm dep}$ and the star-forming gas fraction $f_{\rm sf}$, with the timescales of gas cycle on the scales of individual star-forming regions and their dependency on gas surface density, $\Sgas$. 

The left column shows the values of $\tau_{\rm dep}$ and $f_{\rm sf}$ as a function of $\Sigma_{\rm gas}$, directly measured from our three simulations. The figure shows clear trends with $\Sigma_{\rm gas}$: at higher $\Sigma_{\rm gas}$, depletion times decrease rapidly, approximately as $\taudep \propto {\Sigma_{\rm gas}}^{-1.46}$, while the star-forming fraction increases as $f_{\rm sf} \propto {\Sigma_{\rm gas}}^{0.65}$. Note, that such a strong dependence of $\taudep$ on $\Sgas$ implies a very steep KSR: $\SSFR \propto \Sgas/\taudep \propto \Sgas^{2.46}$. As we showed in Section~\ref{sec:sims-cycle}, such a steep slope results from the contribution of ionized gas and optically thick HI, while the relations for molecular gas and warm, optically thin H{\sc i} produced in our simulations are shallower, consistent with observations (see Figure~\ref{fig:combined_KSR}). 

It is also important to note that in Figure~\ref{fig:model_all}, all three of our simulated galaxies exhibit similar trends, differing only in the range of $\Sigma_{\rm gas}$ they sample. This result suggests that, despite drastic variations in global $\tau_{\rm dep}$ and $f_{\rm sf}$ averaged over entire galaxies, their values on kiloparsec scales are set by the local environment, which in turn is well described by $\Sigma_{\rm gas}$, and not by the global properties of the galaxy (e.g., total mass, gas fraction, size, etc.). Consequently, the global $\tau_{\rm dep}$ and $f_{\rm sf}$ vary due to the differences in the representation of the environmental dependence of $\Sigma_{\rm gas}$: a dwarf galaxy has longer global $\tau_{\rm dep}$ as it only reaches relatively low $\Sigma_{\rm gas}$, while in a gas-rich galaxy, global $\tau_{\rm dep}$ is short due to the high $\Sigma_{\rm gas}$ reached. For this reason, the goal of our gas-cycling framework is to explain the trends of $\tau_{\rm dep}$ and $f_{\rm sf}$ with $\Sigma_{\rm gas}$ by connecting them with the environment-dependent timescales of gas evolution on small scales.

In the following subsections, we investigate the dependence of the ISM gas cycle timescales ($\taup$, $\taum$, and $\taustar$) on the environment and how it can explain the above trends, with the second and third columns of Figure \ref{fig:model_all} summarizing our results. The second column shows the trends of $\tau_{\rm dep}$ and $f_{\rm sf}$ predicted by our framework, where we measure the relevant timescales of the small-scale gas cycle ($\taup$, $\taum$, and $\taustar$) and substitute them in Equations~\ref{eq:fsf}--\ref{eq:Nc}. The trends predicted by our framework are close to those seen in the simulations, indicating that the framework successfully captures the link between kiloparsec-scale star-forming properties and the small-scale gas cycle.

Furthermore, based on our results, we also develop a simple physically motivated model for these timescales (also described below and summarized in Section~\ref{sec:model_summary}), which enables one to connect these timescales with the properties of the ISM on kpc scales, such as $\Sigma_{\rm gas}$, disk thickness, and the average turbulent velocity dispersion. As shown in the right column of Figure \ref{fig:model_all}, our simple model reproduces the trends remarkably well, indicating that it captures the main processes driving the dependence of $\tau_{\rm dep}$ and $f_{\rm sf}$ on $\Sigma_{\rm gas}$ in our simulations. One noticeable exception is that the $f_{\rm sf}(\Sgas)$ for $L_*$ galaxy slightly deviates from the fitted trend at the low-$\Sigma_{\rm gas}$ end (bottom right panel), which can be corrected by accounting for the more complex dependence of the removal timescale $\taum$ on $\Sgas$ than the one adopted in our simple model (see the purple dashed line and Section~\ref{sec:tau_minus} for details).

\subsection{The timescales of the ISM gas cycle} \label{sec:timescales}

\begin{figure*}
    \centering
    \includegraphics[width=\textwidth]{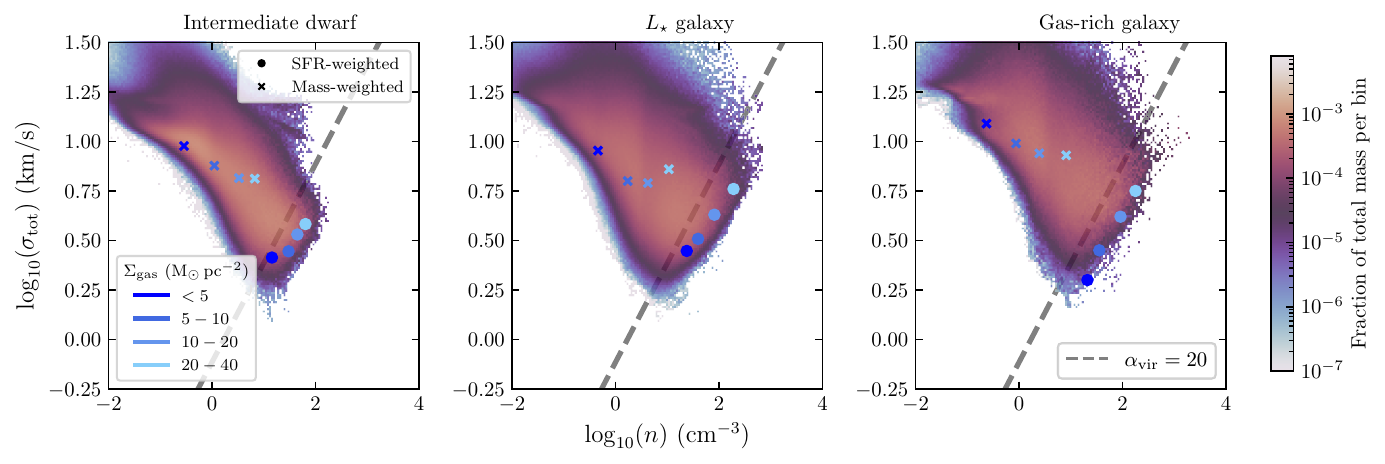}
    \caption{Distribution of the total velocity dispersion, $\sigma_{\rm tot}$, and gas density, $n$, for all three galaxy simulations. The color map shows the fraction of total gas mass per bin, while colored points indicate the median SFR-weighted (dots) and mass-weighted (crosses) values for gas in a given range of surface density, $\Sigma_{\rm gas}$, as labeled. The dashed line corresponds to $\alphvir = 20$, the value used to identify star-forming gas. In all cases, star-forming gas is situated close to this $\alphvir$ value, and the SFR-weighted medians move along the constant $\alphvir$ line as $\Sigma_{\rm gas}$ increases, leading to a rise of both $n$ and $\sigma_{\rm tot}$. These trends directly influence the key timescales of the star formation cycle: higher densities shorten the free-fall time and thus reduce the star formation timescale, $\taustar$, while the accompanying changes in density and turbulent support modify both the supply, $\taup$, and the removal timescale, $\taum$ (see Sections \ref{sec:timescales}--\ref{sec:tau_minus} for a detailed discussion).}
    \label{fig:2d_hist}
\end{figure*}

Figure \ref{fig:model_timescales} shows the gas cycling timescales measured in our simulations (see Sections~\ref{sec:model_overview} and \ref{sec:sims-cycle}) and their dependence on $\Sigma_{\rm gas}$. The local depletion time $\taustar$ and the supply time $\taup$ of star-forming gas are relatively long, $\taup \sim 5\text{--}30$ Myr and $\taustar \sim 200\text{--}2000$ Myr, and decrease with $\Sigma_{\rm gas}$, reflecting shorter dynamical timescales and more efficient local star formation in denser ISM. In contrast, the dispersal time of star-forming gas is very short, $\taum \lesssim 1$ Myr, and only weakly increases with $\Sigma_{\rm gas}$. This weak trend suggests that dispersal processes (stellar feedback, turbulence, and shear) are set by the local properties of star-forming gas that change only weakly with $\Sigma_{\rm gas}$.

As the ability of gas to form stars depends on its local density, $n$, and total velocity dispersion, $\sigma_{\rm tot}$, it is the evolution of these properties that determines the trends of $\taustar$, $\taup$, and $\taum$ with $\Sigma_{\rm gas}$ (recall Figure~\ref{fig:illust}). To elucidate the origin of these trends, Figure \ref{fig:2d_hist} shows the distribution of gas in the $(n, \sigma_{\rm tot})$ plane and the dependence of the average values on $\Sigma_{\rm gas}$ (marked with points). Each point marks the gas mass-weighted (cross) or SFR-weighted (circle) median value in a given range of gas surface density, $\Sigma_{\rm gas}$. The dashed line shows the $\alphvir = 20$ line, corresponding to $\eff \sim 0.002$, which we use to define the star-forming state, i.e., the region under the dashed line in the plot. 

The characteristic timescales of the gas cycle change with $\Sigma_{\rm gas}$ because the average properties of the star-forming and non-star-forming ISM change. As $\Sigma_{\rm gas}$ increases, the SFR-weighted medians---which are representative of the star-forming state---shift along a constant $\alphvir$ dependence toward simultaneously higher densities and larger total velocity dispersions, as gas resides deeper within dense, star-forming regions. The mass-weighted medians---representative of the average, non-star-forming ISM state---also move systematically in the $(\sigma_{\rm tot},\,n)$ plane with increasing $\Sigma_{\rm gas}$, though their progression is different than that of the SFR-weighted medians. As the ambient gas becomes denser, a larger fraction of it cools into the cold and dense state (i.e., the cold neutral medium, CNM, which occupies the part of the $(n, \sigma_{\rm tot})$ plane close to the dashed line), reducing the relative contribution from the average ISM with $n \sim 1\ {\rm cm^{-3}}$ and $T \sim 10^4$ K (or $\sigma_{\rm tot} \sim 10\ {\rm km\,s^{-1}}$; warm neutral medium, WNM). Because the velocity dispersion $\sigma_{\rm tot}$ (a sum of the sound speed and subgrid turbulent support components, see Equation~\ref{eq:sigmatot}) in the WNM is dominated by its relatively large thermal sound speed, shifting mass from WNM to CNM drives the mass-weighted $\sigma_{\rm tot}$ downward at fixed $n$. As a result, the fraction of star-forming gas also increases with $\Sigma_{\rm gas}$ (see Figure \ref{fig:model_all}), hence mass-weighted medians increasingly sample cooler, more denser material, causing the points to shift rightward in the $(n,\,\sigma_{\rm tot})$ plane as $\Sigma_{\rm gas}$ increases.

As detailed in the following subsections, these systematic shifts in both the star-forming and non-star-forming states are tied to the change in the evolution timescales. In short, denser environments allow non-star-forming gas to reach star-forming conditions more rapidly, reducing the timescale associated with entering the star-forming state, $\taup$ (Section~\ref{sec:tau_plus}). At the same time, the rise in star-forming density at higher $\Sigma_{\rm gas}$ shortens the freefall times of star-forming gas, driving down $\taustar$ (Section~\ref{sec:tau_star}). In contrast, gas that has already reached these conditions becomes harder to disperse by feedback or turbulent motions, leading to a longer characteristic timescale for leaving the star-forming state, $\taum$, although this trend is rather weak (Section~\ref{sec:tau_minus}). 

\subsection{The supply timescale, $\taup$} \label{sec:tau_plus}

\begin{figure}[!t]
    \centering
    \includegraphics[trim = 3mm 0mm 0mm 6mm, clip, width=0.45\textwidth]{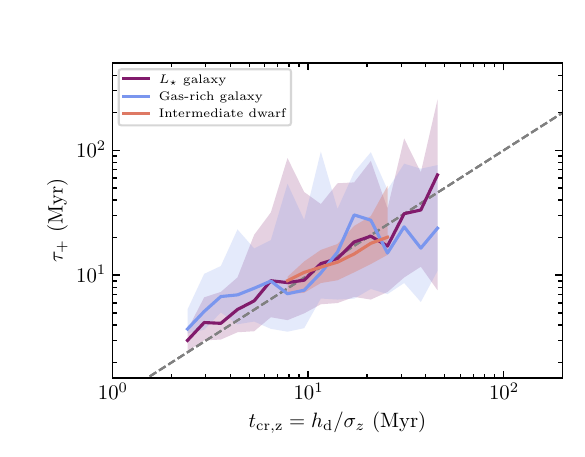}
    \caption{Comparison between the gas supply timescale, $\taup$, and the vertical crossing time $t_{\rm cr,z}$ for our simulated galaxies. The dashed line marks the one-to-one relation. Although the scatter is significant, the median $\taup$ follows the vertical crossing time on average.} 
    \label{fig:tau_plus} 
\end{figure}

\begin{figure}[!t]
    \centering
    \includegraphics[trim = 3mm 0mm 0mm 6mm, clip, width=0.45\textwidth]{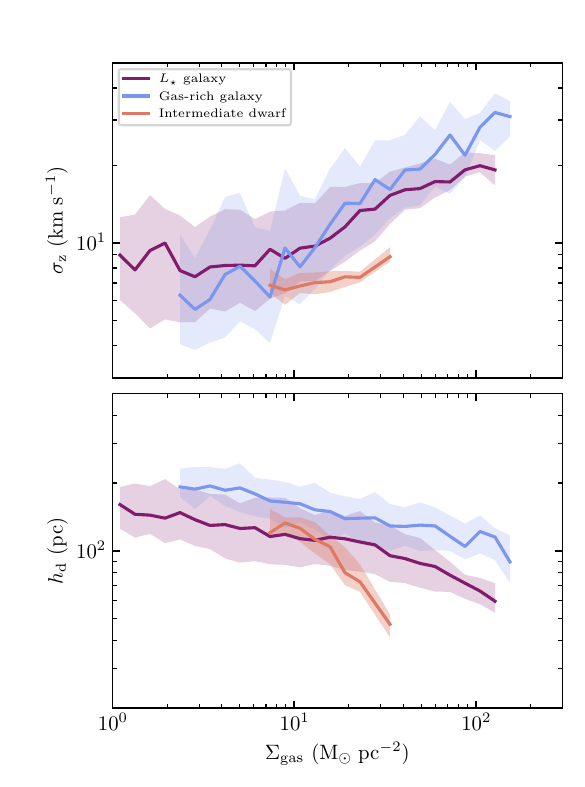}
    \caption{Vertical velocity dispersion $\sigma_{\rm z}$ (top) and local disk scale height $h_{\rm d}$ (bottom) as functions of gas surface density for the three galaxies. The increase in $\sigma_{\rm z}$ and corresponding decrease in $h_{\rm d}$ with $\Sigma_{\rm gas}$ show that denser regions of the disks experience higher vertical pressure and have more compressed vertical structure. These quantities determine the vertical crossing time, $t_{\rm cross,\ z} = h_{\rm d}/\sigma_{\rm z}$, and therefore their trends together explain the trend of $\taup$ with $\Sgas$ (see Figure~\ref{fig:model_timescales}).} 
    \label{fig:tau_plus_comp} 
\end{figure}

First, we examine the timescale associated with the supply of gas into the star-forming state, $\taup$. This timescale characterizes the average time that non-star-forming gas spends before becoming star-forming. Physically, this corresponds to the timescale on which diffuse gas is compressed, cools down, and loses turbulent support sufficiently to reach star-forming conditions, i.e., low $\alphvir \propto \sigma_{\rm tot}^2/\rho$. Consequently, this timescale decreases with increasing $\Sigma_{\rm gas}$ (see Figure \ref{fig:model_timescales}), as the dynamical timescales of gas compression and cooling, as well as the turbulent dissipation times decrease.

The associated timescales can be set by various processes, such as turbulent crossing times, the intervals between spiral arm passages, and fountain outflow recycling \citep[see][]{semenov_physical_2017}. We have investigated several such timescales (see below) and found that the vertical turbulent time, $t_{\rm cross,z} = h_{\rm d}/\sigma_{\rm z}$, describes the values of $\taup$ the best, as shown in Figure \ref{fig:tau_plus}. Here $\sigma_z$ is the mass-weighted vertical velocity dispersion relative to the disk plane due to the resolved vertical motions, which dominate over the subgrid turbulence as they represent the larger, energy-containing scale. Note that this timescale does not include thermal velocity, meaning that the supply time of star-forming gas in our simulations relates to the properties of ISM turbulence rather than the total vertical support of the disk as suggested by a similar connection between depletion times and vertical crossing time that includes both thermal and non-thermal support in marginally stable disks (see Section~\ref{sec:dis}). In our simulations, including the thermal contribution in this timescale worsens the correlation with $\taup$, but it stays within the scatter from the one-to-one relation. 

Such a correlation between $\taup$ and $t_{\rm cross,z}$ can arise as a result of turbulence dissipation, which occurs on the largest turbulent eddy turnover time on the scale of the disk height, $h_{\rm d}$, implying that the star-forming structures can form out of the turbulent flow on a comparable timescale. In Figure \ref{fig:tau_plus_comp}, we show how the two ingredients of $t_{\rm cross,\ z}$, the vertical velocity dispersion $\sigma_{\rm z}$ and the disk scale height $h_{\rm d}$, vary with gas surface density for our different galaxy models. The upper panel shows that $\sigma_{\rm z}$ rises systematically with increasing $\Sigma_{\rm gas}$, reflecting the stronger turbulence in higher-density and more actively star-forming environments. The lower panel shows the mild decrease in $h_{\rm d}$ with $\Sigma_{\rm gas}$, which results from the enhanced gravitational compression in high-surface-density regions. Together, these opposing trends cause the ratio $h_{\rm d}/\sigma_{\rm z}$ to decrease with $\Sigma_{\rm gas}$, leading to shorter $\taup$ at higher $\Sigma_{\rm gas}$.

In addition to the vertical turbulent crossing time, we tested whether $\taup$ correlates with several other characteristic dynamical timescales that describe gas motion on kiloparsec scales. These include the free-fall time at the midplane density, $t_{\rm ff} = \sqrt{3\pi/(32G\rho_{\rm mid})}$, which measures the collapse time set by the total midplane density $\rho_{\rm mid} = \Sigma_{\rm gas}/2h_{\rm d}$, and the in-plane crossing time, $t_{\rm cross,xy} = L / \left[0.5(\sigma_r^2 + \sigma_\theta^2)\right]^{1/2}$, which estimates the time for gas to move across the $L =  1$ kpc-sized region given its radial and azimuthal velocity dispersions, $\sigma_r$ and $\sigma_\theta$. We also considered orbital timescales, $t_{\rm orb} = 2\pi R / \langle v_\theta \rangle$, which trace the rotation period at radius $R$ based on the local mean azimuthal velocity $v_\theta$.
In all cases, we find either significant differences in the normalization or no correlation, suggesting that the transition of gas into the star-forming state is not governed by these timescales.

\subsection{The local depletion time of star-forming gas, $\taustar$} \label{sec:tau_star}

\begin{figure}
    \centering
    \includegraphics[trim = 3mm 0mm 0mm 6mm, clip, width=0.45\textwidth]{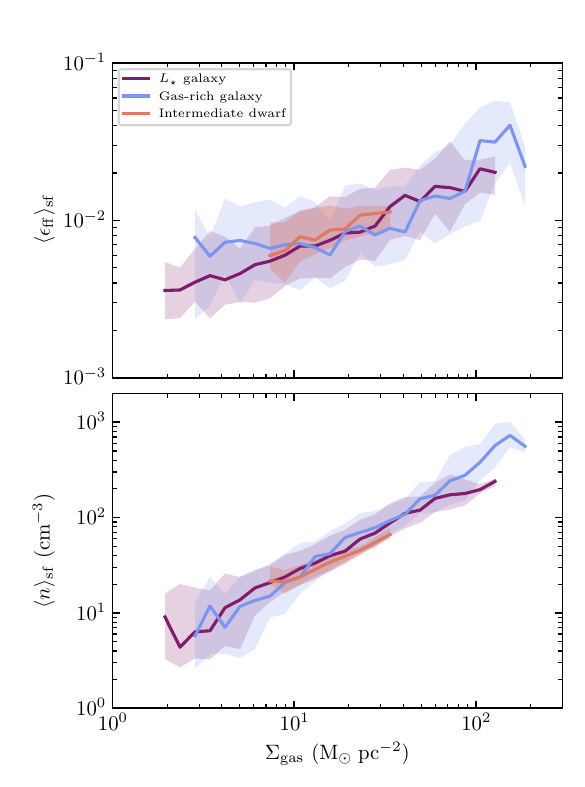}
    \caption{The dependence of the star formation efficiency per free-fall time $\eff$ and average density of the star-forming regions on $\Sgas$. Across all three galaxies, $\eff$ varies only weakly with $\Sigma_{\rm gas}$, whereas $n_{\rm sf}$ increases much more strongly. These quantities determine the local depletion time of star-forming gas, $\taustar = t_{\rm ff}(\langle n \rangle_{\rm sf})/\langle \eff \rangle_{\rm sf}$, and therefore their trends together explain the trend of $\taustar$ with $\Sgas$ (see Figure~\ref{fig:model_timescales}).}
    \label{fig:epsff_n_sf_sigma_gas}
\end{figure}

\begin{figure}
    \centering
    \includegraphics[trim = 3mm 0mm 0mm 6mm, clip, width=0.45\textwidth]{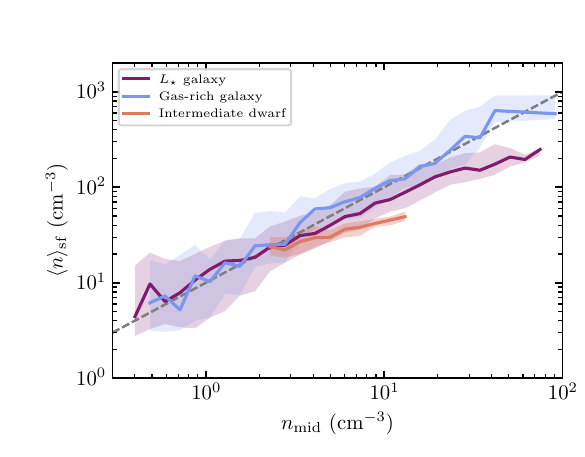}
    \caption{Comparison between the mean density of star-forming gas, $\langle n \rangle_{\rm sf}$, and the midplane density, $n_{\rm mid}$. The dashed line marks the $\langle n \rangle_{\rm sf} = 10\ n_{\rm mid}$ relation. In all three simulations, star-forming gas has densities systematically higher than the midplane average. This separation reflects the fact that star-forming regions occupy the dense, low-$\alphvir$ parts of the ISM, whereas the midplane density traces the total gas distribution averaged over all ISM phases (see crosses and circles in Figure~\ref{fig:2d_hist}).}
    \label{fig:densities}
\end{figure}

The next timescale that exhibits a strong trend with $\Sgas$ in Figure~\ref{fig:model_timescales} is the local depletion time of star-forming gas, $\taustar$. To elucidate the origin of this trend, Figure~\ref{fig:epsff_n_sf_sigma_gas} shows the two ingredients determining $\taustar = \tau_{\rm ff} / \langle \eff \rangle_{\rm sf}$: the local star-formation efficiency per freefall time, $\langle \eff \rangle_{\rm sf}$, and the average density of star-forming gas. For consistency with our star formation prescription (Equation~\ref{eq:rho_sfr}), $\langle \eff \rangle_{\rm sf}$ is weighted by $\rho^{1.5}$ and $\langle n \rangle_{\rm sf}$ is the density corresponding to the average freefall time, inversely averaged by mass, $\tau_{\rm ff} \equiv \langle \tff^{-1} \rangle^{-1}_{\rm sf} \propto \langle n \rangle_{\rm sf}^{-0.5}$ (see Section~\ref{sec:methods}). As the figure shows, both $\langle \eff \rangle_{\rm sf}$ and $\langle n \rangle_{\rm sf}$ exhibit an increasing trend with $\Sgas$, implying that $\taustar \propto \langle n \rangle_{\rm sf}^{-0.5} \langle \eff \rangle_{\rm sf}^{-1}$ becomes shorter as star-forming gas becomes denser and forms stars more efficiently.

The trend of $\langle \eff \rangle_{\rm sf}$ is rather mild: it increases by a factor of $\sim 5$ as $\Sgas$ increases by two orders of magnitude. Such a weak dependence is consistent with the results of \citet{polzin-2024}, who showed that $\langle \eff \rangle_{\rm sf}$ depends only weakly on metallicity and incident radiation field because, locally, $\eff$ in star-forming gas is limited by the dispersal of such gas by feedback on short timescales. The range of $\langle \eff \rangle_{\rm sf}$ variation in Figure~\ref{fig:epsff_n_sf_sigma_gas} is a factor of $\sim 2$ larger than that reported by \citet{polzin-2024} because of the wider range of $\Sgas$ probed in the $L_\star$ and gas-rich galaxy simulations, compared to the intermediate dwarf initial conditions used in that paper.

While $\eff$ changes mildly, the average density of star-forming gas, $\langle n \rangle_{\rm sf}$, increases almost linearly with $\Sgas$. This increase in $\langle n \rangle_{\rm sf}$ with $\Sgas$ reflects the change of the distribution of density and velocity dispersion at different $\Sgas$ as shown in Figure~\ref{fig:2d_hist}: as $\Sgas$ increases, the SFR-weighted average density systematically increases. The increasing trend of $\langle n \rangle_{\rm sf}$ in Figures~\ref{fig:2d_hist} and \ref{fig:epsff_n_sf_sigma_gas} together with the increasing star-forming mass fraction ($\fsf$; see Figure~\ref{fig:model_all}) suggest that at $\Sgas$, a larger fraction of gas participates in star formation and its typical densities become higher, thereby leading to shorter local depletion timescales. 

Quantitatively, the relation between $\Sgas$ and $\langle n \rangle_{\rm sf}$ depends on the detailed distribution of $n$ and $\sigma_{\rm tot}$ on the sub-kpc scale, resulting in the above behavior. The fact that $\langle n \rangle_{\rm sf}$ scales almost linearly with $\Sgas$ and that the disk thickness $h_{\rm d}$ changes only modesty (Figure~\ref{fig:tau_plus_comp}) suggests that the midplane density of gas, $\rho_{\rm mid} = \Sgas/2 h_{\rm d}$, can be used to approximate this trend with the normalization dependent on the details of the $n$--$\sigma_{\rm tot}$ distribution. Indeed, as Figure~\ref{fig:densities} shows, the density of star-forming gas in our simulations is reasonably well approximated by $\langle n \rangle_{\rm sf} \approx 10\,n_{\rm mid}$, which we use to model the dependence of $\taustar$ on $\Sgas$ below. 

\subsection{The removal timescale, $\taum$} \label{sec:tau_minus}

\begin{figure*}
    \centering
    \includegraphics[width=0.95\textwidth]{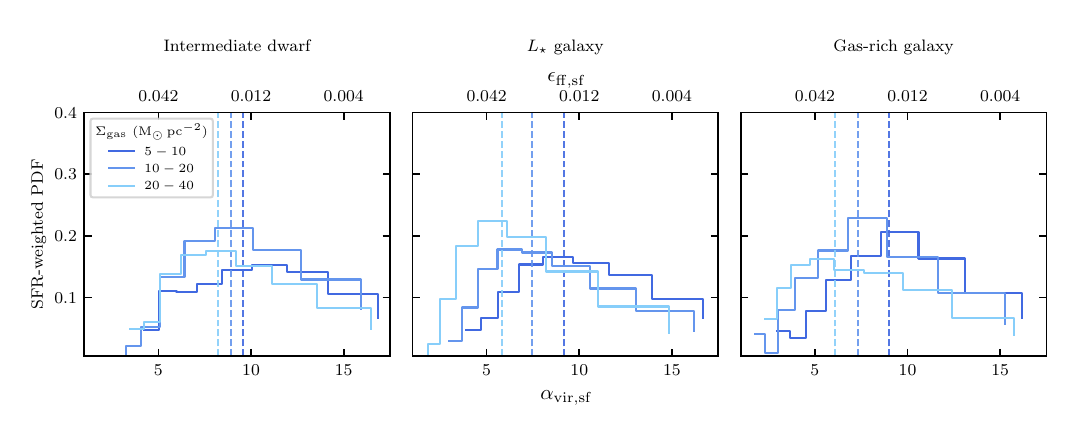}
    \caption{SFR-weighted probability distribution functions (PDFs) of the virial parameter, $\alphvir$, for the three simulated galaxy environments (for gas with $\alphvir < \alpha_{\rm vir, sf} = 20$). Each panel shows the distribution in different ranges of gas surface density, $\Sigma_{\rm gas}$, as labeled. Vertical dashed lines mark the SFR-weighted median values for each $\Sigma_{\rm gas}$ bin. The upper axis shows the corresponding star formation efficiency per free-fall time, $\eff$. Higher gas surface densities correspond to lower median $\alphvir$ and thus higher $\eff$, consistent with more actively star-forming regions in denser environments.}
    \label{fig:PDF_alpha_vir}
\end{figure*}

\begin{figure}
    \centering
    \includegraphics[trim = 3mm 0mm 0mm 6mm, clip, width=0.45\textwidth]{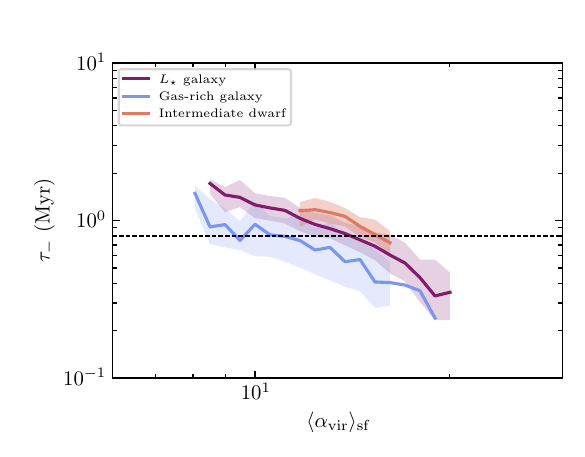}
    \caption{The relation between the removal timescale $\taum$ and average $\alphvir$ for all three galaxies considered. Increasing $\alphvir$ leads to lower values of $\taum$ as less actively star-forming gas, residing in more diffuse environments, can be dispersed more easily, explaining the mild decrease of $\taum$ at lower $\Sigma_{\rm gas}$ in Figure \ref{fig:model_timescales}. Considering the modest variation in the median values of $\taum$, in our model, we adopt a value of $\taum \approx 0.8$ Myr (black dashed line).}
    \label{fig:tau_m_alpha_vir}
\end{figure}

Finally, the timescale of star-forming gas dispersal, $\taum$, is the shortest of the key gas cycle timescales shown in Figure~\ref{fig:model_timescales} and exhibits the weakest trend with $\Sgas$. In the framework of Section~\ref{sec:model_overview}, this timescale is critically important: its short values, $\taustar \lesssim 1$ Myr, explains why the mass fraction of star-forming gas is low $\fsf \sim \taum/(\taup+\taum) \sim \taum/\taup$, and why the star formation is globally inefficient. The latter arises because the number of star formation cycles, $\Nc \sim \taustar/\taum$ is large, which in turn leads to a long global depletion time, much longer than the ISM dynamical timescales, on which star-forming gas forms, $\tau_{\rm dep} \sim \Nc \taup + \taustar \gg \taup$. 

Just like $\taup$ and $\taustar$, the weak increasing trend of $\taum$ with $\Sgas$ in Figure~\ref{fig:model_timescales} can be explained by the change of the gas distribution in the $(n,\,\sigma_{\rm tot})$ space, as shown in Figure~\ref{fig:2d_hist}. As $\Sgas$ increases, gas moves approximately along the constant $\alphvir = 20$ boundary, gradually deviating toward higher densities and lower $\sigma_{\rm tot}$ (see the colored circles in the figure). This deviation implies that the average $\alphvir$ of star-forming gas becomes lower, which is explicitly shown in Figure~\ref{fig:PDF_alpha_vir}. Such lower values of $\alphvir$ indicate that it takes longer for stellar feedback and dynamical processes to inject enough turbulence (increase $\sigma_{\rm tot}$) and/or physically disperse the gas (decrease $n$) to render it non-star-forming, yielding longer $\taum$.

This trend of $\taum$ with the average $\alphvir$ of star-forming gas is shown in Figure \ref{fig:tau_m_alpha_vir}. Across the wide range of surface densities and morphologies examined, however, the variation in the median $\taum$ values remains relatively modest. Therefore, for the purposes of our analytic model, we adopt a representative value of $\taum \approx 0.8$ Myr (grey dashed line), which captures the typical removal timescale across all three simulated galaxies. 
Accounting for the trend of $\taum$ with $\Sgas$ only marginally improves the agreement of the model with simulation results, in particular the fraction of star-forming gas at low $\Sgas$ (see the dashed purple line in the right panel of Figure~\ref{fig:model_all}).

\subsection{Summary of the introduced model} \label{sec:model_summary}

In summary, the behaviour of the three characteristic timescales in Figure~\ref{fig:model_timescales} can be understood as a consequence of the conditions in the turbulent, multiphase ISM that the gas cycles through, which can be described by a simple model summarized here. For reference, this model is shown with the dashed lines in Figure~\ref{fig:model_timescales}, and despite its simplicity, the model captures the behavior of the key timescales remarkably well. Consequently, the model can explain the general trends of the key star-forming properties, total depletion time and star-forming mass fraction (Figure~\ref{fig:model_all}, right column). 

Our simulation results suggest that the supply timescale $\taup$ can be approximated by the vertical turbulent crossing time of the disk (Figure \ref{fig:tau_plus}):

\begin{equation}
    \taup \approx t_{\rm cross, z} = \frac{h_{\rm d}}{\sigma_{\rm z}}.
\end{equation}
Such a parametrization is especially convenient, as both the scale-height $h_{\rm d}$ and vertical velocity dispersion $\sigma_{\rm z}$ can be tied to the kiloparsec scale $\Sgas$ by assuming that the galactic disk is in a marginally stable state. In this picture, $h_{\rm d}$ is set by the level of turbulence ($\sigma_{\rm z}$), driven by stellar feedback, which in turn depends on $h_{\rm d}$ due to the dependence of the average volume densities of gas at fixed $\Sgas$ \citep[e.g.,][]{ostriker_maximally_2011,faucher-giguere_feedback-regulated_2013,ostriker_pressure-regulated_2022}.

The depletion time of the star-forming gas, $\taustar$, decreases with $\Sgas$ as both the density of star-forming and its efficiency increase with $\Sgas$ (Figure~\ref{fig:epsff_n_sf_sigma_gas}). However, given that the trend of $\eff$ is mild, we find that assuming a constant $\eff \sim 0.01$ is sufficient to capture the behavior of $\taustar$. Since the star-forming regions are typically an order of magnitude denser than the midplane gas ($\rho_{\rm mid} = \Sigma_{\rm gas}/2h_{\rm d}$; see Figure \ref{fig:densities}), this timescale becomes
\begin{equation}
    \taustar = \frac{t_{\rm ff}(\rho_{\rm sf})}{\eff} \approx \frac{t_{\rm ff}(10\, \rho_{\rm mid})}{0.01}.
\end{equation}

Finally, the removal timescale $\taum$ traces how long gas remains in the star-forming state before being dispersed by stellar feedback and other dynamical processes. Because the median $\taum$ varies only weakly across surface densities and galaxy types, we adopt a representative value of

\begin{equation}
    \taum \approx 0.8\, \rm Myr,
\end{equation}
and, in Section~\ref{sec:dis:taum-variation}, investigate the effects of its variations.
Note that, although this timescale is closely related to the typical lifetimes of star-forming regions and giant molecular clouds, the lifetimes of such objects as coherent structures can be longer than the above value of $\taum$ (e.g., a few local freefall times, or a few to a few tens of Myr, in \citealt{kim-2022}). This is because $\taum$ represents the Lagrangian lifetime of a gas parcel in a star-forming state, which can be shorter than the lifetime of a coherent star-forming region when gas continuously cycles through such a region \citep{jeffreson-2024}.

\section{Discussion}\label{sec:dis}

\subsection{Comparison to other models}

In this section, we compare our results with other models proposed in the literature that aim to explain the origin of gas depletion times and star-forming mass fractions, as well as their environmental dependence. These models can be broadly divided into several classes based on the underlying description of the gas cycle on the sub-kpc scale.

\textbf{Timescale-Based Gas Cycling.} One class of such models explicitly describes the gas cycle in terms of the timescales on which star-forming gas is created and dispersed, together with efficiency parameters that then quantify star formation over a given timescale (analogous to $\eff$). In such a framework, the kpc-scale depletion time $\taudep$ is directly tied to these quantities \citep[e.g.,][]{madore10,elmegreen15,elmegreen18,burkert_bathtub_2017,maclow-2017}. Methodologically, this approach is similar to our general framework, which enables direct comparison between our simulation results and the underlying assumptions about the timescales adopted in such models \citep[see][for a more detailed comparison]{semenov19}.

A common assumption in these models is that the star-forming gas supply time (our $\taup$) is set by a characteristic gravitational timescale of the average ISM. This timescale is often assumed to be the freefall time evaluated for some characteristic density, e.g., the midplane or volume-averaged gas density \citep[$t_{\rm ff,ISM}$;][]{madore10,elmegreen15,elmegreen18,burkert_bathtub_2017}. We have compared the supply times measured in our simulations with the freefall times at midplane densities and found that, although the two do correlate, $\taup$ are typically a factor of $\sim 3\text{--}5$ times shorter. 

The dispersal times (our $\taum$) in such models are often assumed to be of the order of local freefall time of the star-forming gas, ($t_{\rm ff,sf}$; e.g., the density of molecular transition, $t_{\rm ff,sf} \sim t_{\rm ff,H_2}$, in \citealt{elmegreen15,elmegreen18} or even higher-density self-gravitating molecular state, $t_{\rm ff,sf} \sim t_{\rm ff,dense}$, in \citealt{burkert_bathtub_2017}). These timescales are shorter than the freefall time at the average ISM density, $t_{\rm ff,ISM}$, because gas in star-forming state has significantly higher densities. As a result, only a small fraction of the ISM gas is being actively star-forming: $\fsf \sim \taum/(\taum+\taup) \sim t_{\rm ff,sf}/(t_{\rm ff,sf}+t_{\rm ff,ISM}) \sim t_{\rm ff,sf}/t_{\rm ff,ISM}$ (see Equation~\ref{eq:fsf}). Note that these $t_{\rm ff,sf}$ timescales are independent of the kpc-scale $\Sgas$, as they are fixed by the constant density characterizing the specific ISM conditions in these models (i.e., the H$_2$ formation threshold in \citealt{elmegreen15} and the density defining the ``dense'' state in \citealt{burkert_bathtub_2017}). Although a constant $\taum$ is consistent with our simple model (Section~\ref{sec:model_summary}), the very short value that we find, $\taum \sim 0.8$ Myr, would correspond to $\tff$ at $\rho \sim 4000\ m_{\rm p}\,{\rm cm^{-3}}$. This is significantly higher than the gas densities resolved in our simulations (see Figures~\ref{fig:2d_hist} and \ref{fig:epsff_n_sf_sigma_gas}), indicating that this timescale is not set by a local freefall time. Moreover, our results show that $\taum$ does weakly depend on $\Sgas$ and therefore cannot be set by a fixed density threshold alone (see Figure~\ref{fig:epsff_n_sf_sigma_gas} and the discussion in \citealt{semenov19}). We also note that the Lagrangian lifetime of star-forming gas, $\taum$, can be significantly shorter than the lifetime of star-forming regions as coherent structures, provided that gas continuously cycles through them \citep{jeffreson-2024}.

Finally, for comparisons with such models, it is important to distinguish between the local star-formation efficiencies of the actively star-forming gas (e.g., $\eff$ in our model, which sets $\taustar$), and the total star formation efficiency per cycle, $\epsilon_{\rm int}$. In our model, the latter determines the total number of cycles required to deplete all gas, $\Nc \sim 1/\epsilon_{\rm int}$, and can be connected to $\taustar$ via the lifetime of star-forming gas. Assuming $\taum \ll \taustar$, this yields $\epsilon_{\rm int} \sim \taum/\taustar$ (see also Equation~\ref{eq:Nc}). \citet{madore10} assumes that the total $\epsilon_{\rm int}$ is constant, which contradicts our results: we find that $\epsilon_{\rm int} \sim \Nc^{-1}$ strongly decreases at low $\Sgas$, as $\taustar$ becomes longer with only a moderate change in $\taum$. In contrast, \citet{elmegreen15} and \citet{burkert_bathtub_2017} assume that the local $\eff$ is fixed, which is more in line with our simple model. However, again, its average value does exhibit a weak trend with $\Sgas$ (see Figure~\ref{fig:epsff_n_sf_sigma_gas}), implying the picture is more nuanced. As pointed out by \citet{polzin-2024}, such near-universality of $\eff$ can be explained by the quick dispersal of high-$\eff$ regions by feedback.

\textbf{Effective ISM Models.} Another class of models, broadly described as effective ISM models \citep[e.g.,][]{yepes97,sh03,braun12a}, formulates the ISM gas cycle in terms of a set of differential equations describing mass and energy conservation. In these frameworks, star-forming and non-star-forming states are represented by mass and energy reservoirs, while various formation and destruction processes---such as cooling, heating, and evaporation by supernovae---are described by the specific terms in these equations. All these processes depend on the kpc-scale conditions. Solving the resulting system of equations (either with or without assumptions about steady states for different components), enables one to derive the dependencies of the total SFR and effective ISM pressure on the global properties of the system, such as the average gas density and $\Sgas$. The free parameters of the model can then be calibrated against observations, e.g., the KSR. Conceptually, the general idea behind the effective ISM models is equivalent to the gas-cycling framework adopted in this work, as different formation and destruction terms in such models can be explicitly redefined in terms of the characteristic timescales (Equations~\ref{eq:f_p}--\ref{eq:f_m}). Consequently, our simulation results can be used to inform the modeling of the specific sink and source terms in such models. 

For example, in \citet{yepes97} and \citet{sh03} the balance between star-forming and non-star-forming states (in these models, the ``cold'' and ``hot'' phases, respectively) is mainly driven by the competition between radiative cooling (hot $\to$ cold) and evaporation by supernovae (cold $\to$ hot; modeled following \citealt{mckee77}). Star formation inside the ``cold'' ISM is then parametrized either by a fixed timescale $\taustar$ \citep{yepes97} or a fixed efficiency analogous to $\eff$ \citep{sh03}. 
Our results show that ISM turbulence plays a significant role in the formation of star-forming gas ($\taup$; Figure~\ref{fig:tau_plus_comp}) and setting local depletion times ($\taustar$; Figures~\ref{fig:2d_hist} and \ref{fig:epsff_n_sf_sigma_gas}), warranting future detailed exploration of ISM models that explicitly include the effects of turbulence (see, e.g., \citealt{braun12a} and Kocjan et al., in prep.).

\textbf{Self-Regulated Disks.} The next class of models does not explicitly assume characteristic timescales or rates, but instead relies on the assumption of self-regulation in marginally stable galactic disks \citep[e.g.,][]{ostriker10,ostriker_maximally_2011,faucher-giguere_feedback-regulated_2013,hayward17,ostriker_pressure-regulated_2022}. As an example of such marginally stable disk models, the pressure-regulated feedback-modulated \citep[PRFM;][]{ostriker10,ostriker_maximally_2011,ostriker_pressure-regulated_2022} ISM model predicts that the total gas depletion time in such disks can be described by an integral efficiency per vertical crossing timescale, $\taudep \sim t_{\rm ver}/\epsilon_{\rm ver}$, where 
$t_{\rm ver}$ is analogous to our $t_{\rm cross,z} = h_{\rm d}/\sigma_z$, except that it also includes contributions from all thermal and non-thermal forms of support in the effective $\sigma$, while $\epsilon_{\rm ver}$ is given by the ratio of the so-called feedback yield---the proportionality constant between the SFR surface density and the ISM pressure in their model---and the effective velocity dispersion (see Section~2.3 in \citealt{ostriker_pressure-regulated_2022}). The model further implies that the star-forming mass fraction, $\fsf$, and the local conditions setting its depletion time, $\taustar$, would adjust accordingly to satisfy the condition imposed by the global dynamical equilibrium, ensuring $\fsf = \taustar/\taudep$ (see their Section~2.4).

One difference of our model from PRFM is that the vertical crossing time setting the supply time in our model only includes the turbulent contribution. This relation can be interpreted as a consequence of the typical formation timescale of structures (including star-forming regions) being close to the turbulence dissipation time at the largest turbulent eddy turnover timescale. In the PRFM model, in contrast, the vertical crossing time includes all thermal and non-thermal effective velocities as it originates from the total vertical pressure support of the disk. While in our model, using the turbulent component alone provides a better fit to the simulation data, we note that including the thermal velocity does not lead to a substantially different qualitative behavior. In this sense, our findings remain broadly consistent with the PRFM prediction that gas depletion is regulated by the vertical crossing time associated with the total effective velocity dispersion. 

Given that the underlying timescales in the two models are different, the prefactors describing the integral efficiencies of star formation should also be compared with caution. We note, however, that our simulations show a rather strong dependence of $\epsilon \sim 1/\Nc$ on $\Sgas$, implying that the efficiency per cycle increases at higher $\Sgas$. This trend is qualitatively similar to the one predicted in the PRFM model, resulting from the modest decrease in the feedback yield and an increase in turbulent velocities, although the magnitude of the effect can be different due to the above differences in the vertical crossing time definitions.

Finally, our framework does not require the dynamical equilibrium or even the presence of a disk itself. In conditions other than a marginally-stable disk, the timescale $\taup$ can be set by processes not related to vertical crossing time, such as the gas accretion on the galaxy or outflow gas recycling if feedback drives strong galactic outflows instead of coupling efficiently to the ISM as envisioned by the above models. This motivates a detailed exploration of the gas cycling timescales in more extreme, out-of-equilibrium regimes, such as the first galaxies. Addressing this question requires extending the present analysis beyond isolated disk simulations. A systematic exploration in fully cosmological simulations would help determine how the characteristic timescales $\taup$, $\taum$, and $\taustar$ change in the presence of gas accretion, mergers, and large-scale tidal interactions.

\textbf{Density PDF Models.} Finally, some works attribute the changes in the SFR on kpc scale directly to the behavior of gas PDF on sub-kpc scale \citep[e.g.,][]{kravtsov03,renaud12,gnedin14b,kraljic14,kraljic24}. In such models, the increase of kpc-scale SFR at higher $\Sgas$ reflects the increase in the mass fractions of dense star-forming gas, $\fsf$, as the average densities of the gas increase and, in some models, the shape of the PDF itself changes due to the ISM turbulence transition between subsonic and supersonic regimes. In our gas-cycling framework, such a change in gas PDF can be formulated in terms of the change in the characteristic timescales of processes driving the evolution of relevant gas properties (e.g., density, temperature, and velocity dispersion), thereby determining the PDF of these properties. In this sense, our definition of the timescales $\taup$ and $\taum$ can be thought of as integrals over such PDFs as they represent the average times that gas spends in specific parts of the parameter space (star-forming and non-star-forming in our case). Such a connection between the density PDF and evolution timescales has been investigated in the context of individual star-forming regions in \citet{appel23} and more generally isothermal supersonic turbulence in \citet{scannapieco24}.

\subsection{The Sensitivity of Molecular KSR to Model Timescales} \label{sec:dis:taum-variation}

\begin{figure}
    \centering
    \includegraphics[width=0.48\textwidth]{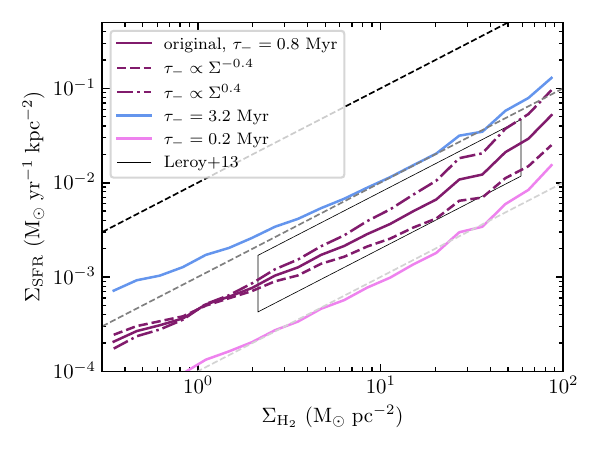}
    \caption{Molecular Kennicutt--Schmidt relation for the $L_\star$ galaxy simulation under different assumptions for the dispersal time of star-forming gas, $\taum$. The molecular gas surface density, $\Smol$, is computed as in Fig.~\ref{fig:combined_KSR}, while the star formation rate surface density is computed as $\SSFR = \Sgas / \taudep$, with $\taudep$ obtained from the model described in Section~\ref{sec:model_summary}. The solid purple curve shows the fiducial model with a constant $\taudep = 0.8$~Myr. Dashed and dash-dotted purple lines correspond to models in which $\taudep$ scales with gas surface density as $\taudep \propto \Sigma^{-0.4}$ and $\taudep \propto \Sigma^{0.4}$, respectively. Blue and pink curves show models with constant depletion times of $\taudep = 3.2$~Myr and $\taudep = 0.2$~Myr. Dashed lines correspond to the constant depletion times $\taudep$ of 0.1 (black), 1 (gray), and 10 Gyr (light gray), for reference. Additionally, we show observational data from \citet{leroy_molecular_2013} in black (the sample-to-sample variation in slope and normalization can be seen in Fig.~\ref{fig:combined_KSR}).
    The changes in $\taum$ lead to systematic effects on the molecular KSR according to Equation~\ref{eq:taudep-taumvar}. From the same equation, the effects of varying $\taup$ and $\taustar$ are similar but opposite in direction.}
    \label{fig:KSR_varied}
\end{figure}

To further illustrate the effects of different processes on the gas depletion times and the observed star formation scaling relations, such as the molecular KSR, we consider the effects of different timescales on these relations. Specifically, we focus on the variation of the gas removal time, $\taum$, for two reasons. First, this timescale parametrizes the effect of stellar feedback, which is one of the key and most uncertain factors regulating depletion times. Second, due to the degeneracies with other timescales, variation of $\taum$ also gives a quantitative idea about the effects of other timescales. Indeed, Equation~\ref{eq:taudep}, can be rewritten as 

\begin{equation}
\label{eq:taudep-taumvar}
    \taudep = \taustar \, \left( 1 + \frac{\taup}{\taum}\right) \approx \taustar \, \frac{\taup}{\taum},
\end{equation}
where we assumed $\taum \ll \taup$, as is the case in our simulations (see Figure~\ref{fig:model_timescales}). From this equation, any changes in $\taum$ are equivalent to the corresponding inverse changes in $\taup$ and $\taustar$.

Figure~\ref{fig:KSR_varied} shows the molecular KSR from our $L_\star$ galaxy simulation, where $\Smol$ is calculated as in Figure~\ref{fig:combined_KSR}, while $\SSFR$ is computed using our simple model for $\taudep$ (Section~\ref{sec:model_summary}) as $\SSFR = \Sgas/\taudep$, with additional variations of $\taum$ value and its dependence of $\Sgas$. 

As the figure shows, the KSR normalization is proportional to $\taum$ (see Equation~\ref{eq:taudep-taumvar}), with shorter $\taum$ (lower $\SSFR$) corresponding to more efficient dispersal of star-forming gas. Similarly, inducing a relatively strong dependence of $\taum$ on $\Sgas$ propagates to the slope of the molecular KSR, although the resulting change in the KSR slope is rather weak, because of the dependence of molecular mass fraction on $\Sgas$. Note, however, that these variations result from variations of $\taudep$ alone, while changes in the feedback model can also affect the abundances of molecular gas, making the effect on the molecular KSR stronger. These results, therefore, should be interpreted as an illustration of the direction of the effects.

As noted above, variations of $\taup$ will affect $\taudep$ similarly to $\taum$, but in the opposite direction. Such variations of the supply time can be driven by variations in the dynamical timescales caused by the differences in the detailed galactic structure and the presence of bars and bulges, and therefore might be contributing to the galaxy-to-galaxy variations of the KSR \citep[see, e.g.,][]{bigiel_star_2008,leroy13}. Studying this effect will require repeating our analysis in multiple simulations representing such environments, which we leave to future work.

\subsection{Extensions of the gas cycling framework to other regimes}

While our results describe star-forming environments characteristic for actively star-forming disks, here we discuss how the interpretation of these timescales can change when we extend our model beyond this regime into more extreme conditions.

For example, if the trends shown in Figure~\ref{fig:model_timescales} continue at higher $\Sgas$, our results suggest that at sufficiently high gas surface densities, the formation time of star-forming gas can become so rapid that most of the ISM gas becomes star-forming: $\taup < \taum$, implying $f_{\rm sf} \sim 1$. In this limit, instead of cycling between states, the entire ISM behaves as a single star-forming region, where the global depletion time would be set by the local processes regulating $\taustar$. Such a transition could induce a qualitative change in the behavior of depletion times at very high densities in high-redshift and starburst galaxies \citep{kennicutt98,genzel10,tacconi17,kennicutt21}. Applying our framework in this regime would require changing the definition of star-forming gas, shifting it to smaller scales and denser environments, for example, by considering the formation and dispersal of dense star-forming cores. For an example of such an application of a similar framework on the scales of individual star-forming regions see \citet{appel_what_2023}.

Another interesting regime is that of early and low-mass galaxies, before the formation of stable disks. For example, \citet{semenov_uv-bright_2025} showed that the ISM gas cycle can undergo a qualitative transition as a result of disk formation. Before disk formation, the ISM is highly turbulent and star formation is highly bursty, with rapid fluctuations in star formation efficiency driven by violent accretion and galaxy mergers. At this stage, these large-scale processes will control the gas evolution on small scales, while the transition to the turbulence driving in the disk environment occurs later, as the galactic disk settles. The behavior of the characteristic timescales ($\taup$, $\taum$, $\taustar$) can strongly deviate from our simple model in such extreme environments, but our framework can still be used to investigate this highly non-equilibrium behavior. We leave a study of the gas cycle in early galaxies using cosmological simulations to future work.

\section{Conclusions} \label{sec:sum}

In this paper, we have investigated the origin of the star formation scaling relations between the total gas depletion time, $\taudep = \Sgas/\SSFR$, the star-forming gas mass fraction, $\fsf = \Sigma_{\rm sf}/\Sgas$, and the gas surface density $\Sgas$ on kiloparsec scales, all of which are the key ingredients of the observed Kennicutt--Schmidt relation. To elucidate these trends, we employed the framework of \citealt{semenov_physical_2017}, which explicitly connects kiloparsec-scale $\taudep$ and $\fsf$ to the timescales of the rapid, continuous ISM gas cycle on the scales of individual star-forming regions. These include the timescales for the formation, dispersal, and local depletion of the star-forming gas: $\taup$, $\taum$, and $\taustar$ (see Section~\ref{sec:model_overview} and Figure~\ref{fig:illust}). We measure these timescales using passive gas tracer particles in a suite of galaxy simulations spanning a range of environments, from a dwarf disk and a Milky Way-mass galaxy to a gas-rich $z \sim 2$ starburst analog (Section~\ref{sec:sims}). In summary, we find the following:

\begin{enumerate}
    \item Our simulated galaxies span different regimes but produce similar trends of $\taudep$ and $\fsf$ with $\Sgas$, differing primarily in the range of $\Sgas$ they sample. This implies that, on kpc scales, both $\taudep$ and $\fsf$ are set by the local environment, which is reasonably well described by $\Sgas$. In contrast, the global differences between galaxies arise from variations in their average $\Sgas$ (Figure~\ref{fig:model_all}; left column).

    \item The timescales of the small-scale ISM gas cycle measured in our simulations, $\taum$, $\taup$ and $\taustar$, also exhibit well-defined trends with $\Sgas$ (Figure~\ref{fig:model_timescales}). Combined with our analytic framework, these trends can explain the dependence of $\taudep$ and $\fsf$ on $\Sgas$ in terms of normalization, slope, and scatter (Figure~\ref{fig:model_all}; middle column). Additionally, based on our results, we have introduced a simple model, which describes the trends of individual timescales with $\Sgas$ (Section~\ref{sec:model_summary}). Despite its simplicity, this model also reproduces the results of our simulations, suggesting that it captures the main processes driving the small-scale gas evolution reasonably well (Figure~\ref{fig:model_all}; right column).

    \item The timescale on which star-forming gas is supplied from the average ISM, $\taup$, ranges from $\sim3$ to 30 Myr and decreases at higher $\Sgas$ (Figure~\ref{fig:model_timescales}). Such values are comparable with the relatively short dynamical timescales driving gas evolution in the ISM. In particular, we find that the vertical turbulent crossing time approximates this timescale across the regimes we have explored here (Figure~\ref{fig:tau_plus}). The variation of the crossing time, in turn, is driven by the increase in the vertical turbulent velocity and the decrease in the disk thickness at high $\Sgas$ (Figure~\ref{fig:tau_plus_comp}).

    \item The timescale on which star-forming gas is converted into stars, $\taustar$, is $\sim200$--$2000$ Myr and, similar to $\taup$, also decreases at higher $\Sgas$ (Figure~\ref{fig:model_timescales}). The decreasing trend of $\taustar$ with $\Sgas$ can be explained by the change in the local properties of star-forming regions: at higher $\Sgas$, these regions become denser and form stars at moderately higher efficiencies per freefall time $\eff$ (Figure~\ref{fig:epsff_n_sf_sigma_gas}). These trends can be approximated by our simple model that assumes a fixed $\eff = 0.01$ and the mean star-forming density a factor of 10 higher than the midplane density (Figure~\ref{fig:densities}).

    \item The dispersal time of star-forming gas, $\taum$, is very short, $\sim 0.4\text{--}1$ Myr and, unlike $\taup$ and $\taustar$, exhibits only a weak increasing trend with $\Sgas$ (Figure~\ref{fig:model_timescales}). This trend can also be attributed to a change in the local properties of star-forming regions; specifically, the decrease of the average virial parameter of star-forming gas at higher $\Sgas$, as the gas becomes closer to being gravitationally bound (Figures~\ref{fig:2d_hist}, \ref{fig:PDF_alpha_vir}, and \ref{fig:tau_m_alpha_vir}). However, we find that neglecting this trend and assuming a fixed $\taum = 0.8$ Myr is sufficient to explain the $\taudep$ and $\fsf$ trends in our simulations.
\end{enumerate}

These results demonstrate the ability of our analytic framework to explain the star formation scaling relations on kiloparsec scales. In the simulations explored here, $\taudep$ and $\fsf$ are set by the competition between the turbulence driven on the galactic disk scale (via $\taup$), the local processes controlling star formation ($\taustar$) and the dispersal of star-forming regions ($\taum$). Note, however, that the balance between these processes and the relative importance of other processes (such as accretion, galaxy mergers, galactic outflow launching, among others) can substantially change in other environments, especially in the absence of well-defined galactic disks. It will therefore be insightful to further investigate the ISM gas cycle in such extreme environments, including the formation of the first galaxies, using cosmological simulations.

\begin{acknowledgments}

We would like to thank Benedikt Diemer, Lars Hernquist, and Andrey Kravtsov for their comments that improved the manuscript. 
ZK acknowledges the support from the Fulbright Graduate Award awarded by the Polish-U.S. Fulbright Commission (Scholarship Agreement No. PL/2023/2/GS).
Support for V.S. was provided by Harvard University through the Institute for Theory and Computation Fellowship. 
The simulations presented in this paper were carried out on the FASRC Cannon cluster supported by the FAS Division of Science Research Computing Group at Harvard University.
Analyses presented in this paper were greatly aided by the following free software packages: {\tt yt} \citep{yt}, {\tt NumPy} \citep{numpy_ndarray}, {\tt SciPy} \citep{scipy}, and {\tt Matplotlib} \citep{matplotlib}. We have also used the Astrophysics Data Service (\href{http://adsabs.harvard.edu/abstract_service.html}{ADS}) and \href{https://arxiv.org}{arXiv} preprint repository extensively during this project and writing of the paper.

\section{contribution}

Using the CRediT (Contribution Roles Taxonomy) system (\url{https://authorservices.wiley.com/author-resources/Journal-Authors/open-access/credit.html}), the main roles of the authors were:\\
\noindent
\textbf{ZK}: conceptualization, methodology, formal analysis, investigation, visualization, writing – original draft.\\
\textbf{VS}: conceptualization, methodology, data curation, resources, supervision, validation, writing – review \& editing.

\end{acknowledgments}

\bibliographystyle{aasjournal}
\bibliography{timescales_paper,vs}

\end{document}